\documentclass[epjCONF]{svjour}
\usepackage{graphicx}
\usepackage[varg]{txfonts} 
\usepackage[latin1]{inputenc}
\usepackage{amsfonts,multicol}
\usepackage[english]{babel}
\renewcommand{\vec}[1]{\mathbf{#1}}
\def\be{\begin{equation}}
\def\ee{\end{equation}}
\def\bea{\begin{eqnarray}}
\def\eea{\end{eqnarray}}
\newcommand{\beq}{\begin{equation}}
\newcommand{\eeq}[1]{\label{#1} \end{equation}} 
 
\session-title{%
19$^{\textnormal{\footnotesize th}}$ International %
IUPAP Conference on Few-Body Problems in Physics%
}
\begin{document}
\title{%
Deeply Bound Kaonic Clusters
}%
\author{%
E. Oset\inst{1}\fnmsep\thanks{\email{oset@ific.uv.es}} 
\and %
V. K. Magas\inst{2} 
\and %
A. Ramos\inst{2}
\and %
S. Hirenzaki\inst{3}
 \and %
J. Yamagata-Sekihara\inst{1,4}
 \and %
A. Martinez Torres\inst{1}
\and %
K. P. Khemchandani\inst{5}
\and %
M. Napsuciale\inst{6}
\and %
L.S. Geng\inst{1}
\and %
D. Gamermann\inst{1}
}
\institute{%
Departamento de F\'{\i}sica Te\'orica and IFIC,
Centro Mixto Universidad de Valencia-CSIC, \\
46100 Burjassot (Valencia), Spain
\and %
Departament d'Estructura i Constituents de la Mat\`eria,
Universitat de Barcelona, \\
Diagonal 647, 08028 Barcelona, Spain
\and %
Department of Physics, Nara Women's University, Nara 630-8506, Japan
\and %
Yukawa Institute for Theoretical Physics, Kyoto University, 
Kyoto 606-8502, Japan
\and %
Centro de F\'isica Te\'orica, Departamento de F\'isica,
 Universidade de Coimbra, P-3004-516 Coimbra, Portugal
\and %
Departamento de F\'{\i}sica, Divisi\'{o}n de Ciencias e Ingenier\'{\i}as,
Universidad de Guanajuato, Campus Le\'{o}n, Lomas del Bosque 103,
Fraccionamiento Lomas del Campestre, 37150, Le\'{o}n, Guanajuato, M\'{e}xico.
}
\abstract{
In this talk I make a review on the theoretical and experimental
situation around the deeply bound kaon clusters, the possible bound kaonic
states of nuclear, rather than atomic nature. At the same time I discuss
novel developments around other kind of bound kaon clusters, which
include states of two mesons and one baryon, with either one or two kaons,
and states of a vector meson and two kaons, which explain naturally the
observed properties of the X(2175) and Y(4260) resonances.
} 
\maketitle
%
%
%
\section{Introduction}
In this talk I will review the present situation around the deeply bound 
kaonic states. The negative kaons experience an attraction in nuclear
matter, which has been proved experimentally by the observation of 
shifts in energy of the kaonic atoms levels. For these kaonic states the
Coulomb interaction and the strong one participate on equal footing, but the
strength of the strong attraction is found to be large enough to
accommodate bound states of nuclear type, where the radius of the kaon is
similar to that of the nucleons. The question arises that at the same time
the kaon finds modes of annihilation in the presence of other nucleons
and the widths of these states can be much larger than the separation
between the levels, in which case the possibility to find peaks
experimentally fades away. This seems to be the case according to
calculations made using chiral dynamical models. Yet, this has not
precluded the experimental search for such states and in several
experiments claims have been made for their finding.  Unfortunately, the
claims were based on misinterpretation of peaks, that devoted works,
simulating the experiments and the reactions taking place there, have
shown could be reproduced in term of conventional, unavoidable reactions
occurring in the experiment. Work follows in different laboratories
looking for other signals and for the lightest of all these clusters, the
$\bar{K} NN$ system. 

    I also devote some time to describe new systems, which have been
studied very recently, and which involve other kind of clusters of kaons.
These states, however, are not controversial. Some of them correspond to
states already known, offering a particular interpretation of their
nature,
and others are predictions for which there could be already some
experimental evidence, but which require more work to be settled. 
  The known states to which I refer are the low lying excited $1/2^+$, 
S=-1 resonances, which appear naturally as bound states or resonances of
systems of two mesons and one baryon in coupled channels, one of the
mesons being a kaon. Another state, which is claimed to be seen in the 
$\gamma p \to K^+ \Lambda$ reaction around 1920 MeV, corresponds to a 
bound
state of $K \bar{K} N$ with the $K$ and $\bar{K}$ coupled to make the 
$f_0(980)$ or $a_0(980)$ resonance.  Finally, two more states recently
found at BABAR and other labs, the X(2175) and Y(4260), are shown to be 
well reproduced as resonances of the $\phi K \bar{K}$ and  
$J/\psi K \bar{K}$ systems respectively. 
  These three body systems have all
been studied with a novel approach to the Faddeev equations in coupled
channels, using chiral unitary dynamics and the on-shell two body
amplitudes, after the useful finding that there is a cancellation of
off-shell parts of the two body amplitudes with the three body amplitudes
originated by the same chiral Lagrangians. This finding is relevant since
it allows one to use empirical amplitudes to solve these Faddeev equations 
without resorting to the use of potentials and their unavoidable off-shell
extrapolation. It also has practical simplifying consequences in the
solution of the Faddeev equations that will be discussed.

 In the next two sections we study the basic $KN$ interaction using chiral
 dynamics and the chiral unitary approach in coupled channels. In the two
 following sections we study the kaons in a nuclear medium and review the
 situation of the deeply bound kaon states. A more detailed study is made
 of one of the reactions claimed to provide evidence of a deeply bound
 kaon cluster from correlated $\Lambda d$ pairs emitted after the 
 absorption of kaons at rest in nuclei. The $\bar{K}NN$ system is reviewed
 in another section and in a further section we study the novel three 
 body systems which lead to the interesting states mentioned above.

\label{SchmidtPL_intro}

\section{Meson-nucleon amplitudes to lowest order}

Following \cite{Eck95,Be95} we write the lowest order chiral Lagrangian,
coupling the octet of pseudoscalar mesons to the octet of $1/2^+$ baryons, as

$$
L_1^{(B)} = < \bar{B} i \gamma^{\mu} \nabla_{\mu} B> - M_B <\bar{B} B> +
$$

\begin{equation}
\frac{1}{2} D <\bar{B} \gamma^{\mu} \gamma_5 \left\{ u_{\mu}, B \right\} >
+ \frac{1}{2} F <\bar{B} \gamma^{\mu} \gamma_5 [u_{\mu}, B]>,
\end{equation}
where the symbol $< \, >$ denotes trace of SU(3) matrices and

\begin{equation}
\begin{array}{l}
\nabla_{\mu} B = \partial_{\mu} B + [\Gamma_{\mu}, B] \\
\Gamma_{\mu} = \frac{1}{2} (u^+ \partial_{\mu} u + u \partial_{\mu} u^+) \\
U = u^2 = {\rm exp} (i \sqrt{2} \Phi / f) \\
u_{\mu} = i u ^+ \partial_{\mu} U u^+
\label{eq:Gamma}
\end{array}
\end{equation}

The SU(3) matrices for the mesons and the baryons are the following

\begin{equation}
\Phi =
\left(
\begin{array}{ccc}
\frac{1}{\sqrt{2}} \pi^0 + \frac{1}{\sqrt{6}} \eta & \pi^+ & K^+ \\
\pi^- & - \frac{1}{\sqrt{2}} \pi^0 + \frac{1}{\sqrt{6}} \eta & K^0 \\
K^- & \bar{K}^0 & - \frac{2}{\sqrt{6}} \eta
\end{array}
\right)
\end{equation}

\begin{equation}
B =
\left(
\begin{array}{ccc}
\frac{1}{\sqrt{2}} \Sigma^0 + \frac{1}{\sqrt{6}} \Lambda &
\Sigma^+ & p \\
\Sigma^- & - \frac{1}{\sqrt{2}} \Sigma^0 + \frac{1}{\sqrt{6}} \Lambda & n \\
\Xi^- & \Xi^0 & - \frac{2}{\sqrt{6}} \Lambda
\end{array}
\right)
\end{equation}

At lowest order in momentum, that we will keep in our study, the interaction
Lagrangian comes from the $\Gamma_{\mu}$ term in the covariant derivative
and we find

\begin{equation}
L_1^{(B)} = < \bar{B} i \gamma^{\mu} \frac{1}{4 f^2}
[(\Phi \partial_{\mu} \Phi - \partial_{\mu} \Phi \Phi) B
- B (\Phi \partial_{\mu} \Phi - \partial_{\mu} \Phi \Phi)] >
\label{eq:Laglow}
\end{equation}

\noindent
which leads to a common structure of the type
$\bar{u} \gamma^u (k_{\mu} + k'_{\mu}) u$ for the different channels, where
$u, \bar{u}$ are the Dirac spinors and $k, k'$ the momenta of the incoming
and outgoing mesons.

We take the $K^- p$ state and all those that couple to it within the chiral
scheme. These states are $\bar{K}^0 n, \pi^0 \Lambda, \pi^0 \Sigma^0,$
$\pi^+ \Sigma^-, \pi^- \Sigma^+, \eta \Lambda, \eta \Sigma^0, K^0 \Xi^0, 
K^+ \Xi^-$. Hence we have
a problem with ten coupled channels. 
The lowest order amplitudes for these channels are easily evaluated from eq.
(5) and are given by

\begin{equation}
V_{i j} = - C_{i j} \frac{1}{4 f^2} \bar{u} (p') \gamma^{\mu} u (p)
(k_{\mu} + k'_{\mu})
\end{equation}

\noindent
were $p, p' (k, k')$ are the initial, final momenta of the baryons (mesons).
Also, for low energies one can safely neglect the spatial components in eq.
(6) and only the $\gamma^0$ component becomes relevant, hence simplifying
eq. (6) which becomes

\begin{equation}
V_{i j} = - C_{i j} \frac{1}{4 f^2} (k^0 + k'^0)
\end{equation}

The matrix $C_{i j}$, which is symmetric, is given in \cite{angels}.

\section{Coupled channels Bethe Salpeter equations}

Following \cite{npa} we write the set of Bethe Salpeter equations in
the $\bar{K} N$ centre of mass frame

\begin{equation}
T_{i j} = V_{i j} + V_{i l} \; G_l \; T_{l j}
\label{eq:BS}
\end{equation}

\noindent
where the indices $i,j$ run over all possible channels. Eqs. (\ref{eq:BS}) are
coupled channels integral equations involving the off shell part of V and T. 
However, by using the N/D method and
neglecting the left hand cut, as done in Quantum Mechanics, one can show that a
very useful form of these equations is possible, by writing the potential and
the T matrix on shell and factorizing them outside the integral involving VGT.  A
different derivation is provided in \cite{juanoset} starting from a potential in
momentum space, which is separable and has a cut off in momentum

\begin{equation}
V(\vec{q},\vec{q'})= V \Theta(\Lambda -|\vec{q}|)~\Theta(\Lambda -|\vec{q'}|)
\end{equation}

Thus the BS equations result into the algebraic matrix equations

\begin{equation}
T = V + V \, G \, T
\end{equation}

\noindent
or equivalently

\begin{equation}
T = [1 - V \, G]^{-1}\, V
\end{equation}

\noindent
with $G$ a diagonal matrix given by

\begin{eqnarray}
G_{l} &=& i \, \int \frac{d^4 q}{(2 \pi)^4} \, \frac{M_l}{E_l
(\vec{q})} \,
\frac{1}{k^0 + p^0 - q^0 - E_l (\vec{q}) + i \epsilon} \,
\frac{1}{q^2 - m^2_l + i \epsilon} \nonumber \\
&=& \int \, \frac{d^3 q}{(2 \pi)^3} \, \frac{1}{2 \omega_l
(q)}
\,
\frac{M_l}{E_l (\vec{q})} \,
\frac{1}{p^0 + k^0 - \omega_l (\vec{q}) - E_l (\vec{q}) + i \epsilon}
\end{eqnarray}

\noindent
which is regularized using either dimensional regularization or a cutoff,
$q_{max}$, and depends on $p^0 + k^0 = \sqrt{s}$ and a subtraction
constant in dimensional regularization or $q_{max}$ in the cutoff method.

It is interesting to note here, because it will be used later on, that the
Lagrangian of eq. (\ref{eq:Laglow}) has kept only two mesons in the expansion
of $\Gamma_{\mu}$ of eq. (\ref{eq:Gamma}). One can keep up to four mesons in the expansion and obtain amplitudes for two mesons one baryon going to also two mesons and one baryon, which will be used in the section devoted to three body states.

\section{Kaons in a nuclear medium}

The model discussed in the former section for the $\bar{K}N$
interaction gives rise to a s-wave $\bar{K}$ self-energy (${\bar
K}=K^-$ or $\bar{K}^0$)
\begin{equation}
\Pi^s_{\bar{K}}(q^0,{\vec q},\rho)=2\int \frac{d^3p}{(2\pi)^3}
n(\vec{p}) \left[ T_{\rm eff}^{\bar{K}
p}(P^0,\vec{P},\rho) +
T_{\rm eff}^{\bar{K} n}(P^0,\vec{P},\rho) \right] \ ,
\label{eq:selfka}
\end{equation}
which is obtained by summing the in-medium ${\bar K}N$
interaction, $T_{\rm eff}^\alpha$ ($\alpha={\bar K} p, {\bar
K}n$), over the nucleons in the Fermi sea. The values
$(q^0,\vec{q}\,)$ stand now for the energy and momentum of the
$\bar{K}$ in the lab frame.
Note that a self-consistent approach is required since one
calculates the ${\bar K}$ self-energy from the effective
interaction $T_{\rm eff}$ which uses ${\bar K}$ propagators which
themselves include the self-energy being calculated.

 The T matrices in eq. (\ref{eq:selfka}) are modified in the medium to take
 into account Pauli blocking on the intermediate nucleons, pionic selfenergies
 and kaon selfenergies of the intermediate states. The p-wave contributions coming from the coupling of the $\bar{K}$ meson
to hyperon particle-nucleon hole $(YN^{-1})$ excitations are also taken into
account.

The method oulined above has been used in  \cite{Koch94,WKW96,Waas97,Lutz98} considering Pauli blocking effects. The work of \cite{angelsself} considers the selfenergy of the pions in addition and the selfenergy of the kaons selfconsistently. The selfconsistency requirement is mandatory in the presence of the nearby $\Lambda(1405)$ resonance and is also implemented in 
\cite{Lutz98,schaffner,galself}, where similar results are obtained. The p-waves have been considered in addition in \cite{Tolos:2006ny}.

\begin{figure*}[!htb]
\centering
\includegraphics[width=0.45\linewidth]{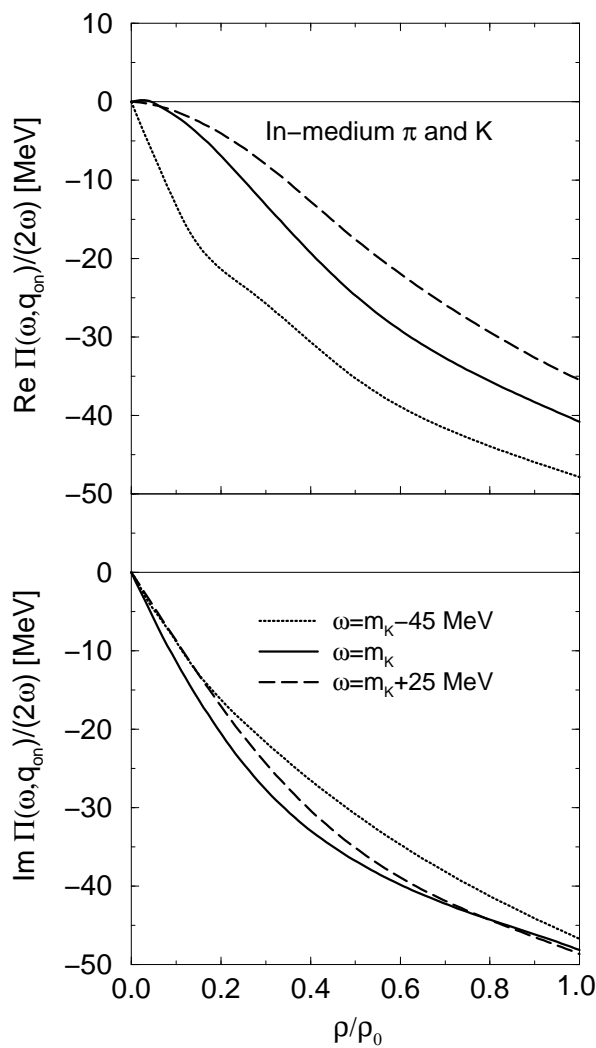}
\caption{
Real (top) and imaginary (bottom) parts of the $K^-$ optical
potential as
a function of density obtained 
from the {\it In-medium pions and kaons} approximation.
Results are shown for three different $K^-$ energies:
$\omega=m_{K}-45$ MeV (dotted lines), $\omega=m_{K}$ (solid lines)
and
$\omega=m_{K}+20$ MeV (dashed lines).
\label{fig:optpotpika}}
\end{figure*}

With this method we obtain a shallow potential which is common to all the chiral
approaches including selfconsistency, and which we show in fig.  \ref{fig:optpotpika}.

\begin{figure*}[!htb] 
\centering
\includegraphics[width=0.55\linewidth]{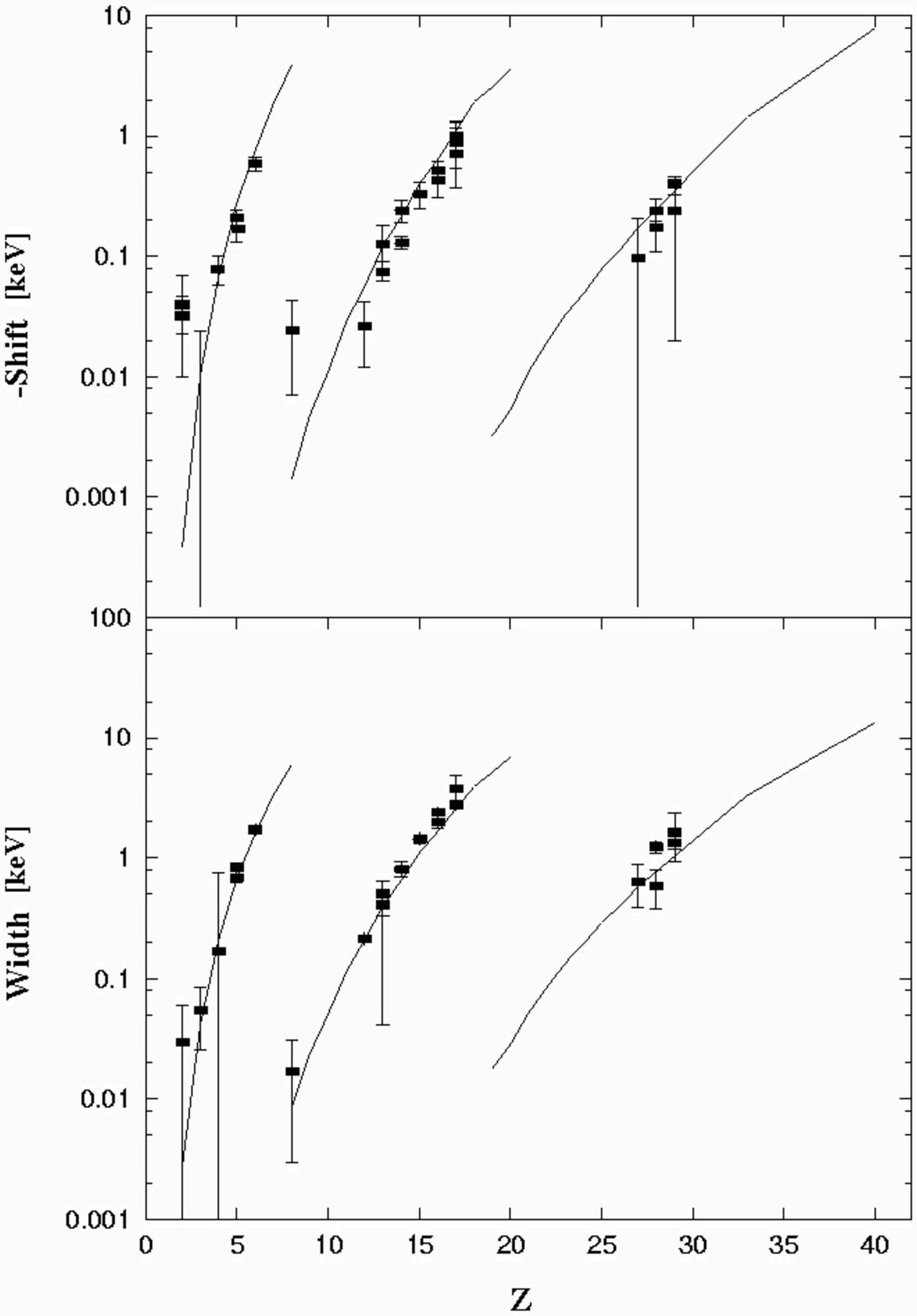}
\caption{Energy shifts and widths of kaonic atom states
(from
 \cite{okumura}). . From left to right the states correspond
to the 2p,3d and 4f atomic levels.}
\label{fig:katom}
\end{figure*}

  It is interesting to note that with this potential one could get a good
reproduction of shifts and widths of kaonic atoms \cite{okumura}, as one can see in fig. \ref{fig:katom}.

A fit to the global set of kaonic atoms data was carried out in \cite{baca},
were it was concluded that a best fit potential could be achieved with a moderate
increase by about 20 \% of the theoretical potential of \cite{angelsself}.

   It is most opportune to mention here that the discrepancies of the theory 
with experiment for the shift of the $^4He$ data in fig. \ref{fig:katom} have been resolved recently,
thanks to a technological breakthrough in the work of \cite{hayano}. It is also
interesting to note that this discrepancy has been used in the past to justify
that the kaon nucleus potential had to have much larger strength \cite{akaishi:2002bg,akainew}. The
new  experimental findings rule out these superstrong potentials with as much as
600 MeV attraction at the center of the nucleus.  

\section{Deeply Bound Kaon Atoms}
   The potential of \cite{angelsself} is sufficiently strong to produce deeply
bound kaonic states. It produces indeed a 1s deeply bound state around 30 MeV.
It produces also a p state bound by 10 MeV.  The problem in both cases is that
the states have a width of about 100 MeV. This is much bigger than the
separation between the levels and precludes the  experimental observation of
peaks.  This is the situation of the deeply bound kaon atoms in the context of
chiral theories, which have proved to be highly accurate to deal
with meson baryon and meson nucleus interaction.

  The issue of the kaon interaction in the nucleus has attracted much
attention in past years. Although from the study of kaon atoms one knows that
the $K^-$-nucleus potential is attractive \cite{friedman-gal}, the discussion
centers on how attractive the potential is and if it can accommodate deeply
bound kaon atoms (kaonic nuclei), which could be observed in direct reactions.
 A sufficiently large attraction could even make possible the existence of
 kaon condensates in  nuclei, which has been suggested in \cite{Kaplan:1986yq}.
The list of papers where strongly attractive potentials are used is long
\cite{gal1,gal2,gal3,gal4,gal,muto,amigo1,amigo2}, 
providing as much as 200 MeV attraction at normal nuclear matter. More moderate
attraction is found in similar works done in 
\cite{Zhong:2004wa,Zhong:2006hd,Dang:2007ai}. Yet, all modern potentials
based on underlying chiral dynamics of the kaon-nucleon interaction 
\cite{Lutz98,angelsself,schaffner,galself,Tolos:2006ny} lead to
moderate potentials of the order of 60 MeV attraction at nuclear matter density.
They also
have a large imaginary part which makes the width of the deeply bound states
much larger than the energy separation between the levels, which would rule out
the  experimental observation of peaks. 

The opposite extreme is represented by some highly 
    attractive phenomenological potentials with about
600 MeV strength in the center of the nucleus \cite{akaishi:2002bg,akainew}.  These
potentials, leading to compressed
nuclear matter of ten times nuclear matter density, met criticisms from 
\cite{toki}  and more 
recently from  \cite{Hyodo:2007jq}, which were rebutted in 
\cite{akanuc} and followed by further argumentation in  \cite{Oset:2007vu} and
 \cite{npangels}. More recently the lightest K-nuclear system of $\bar{K}NN$
 has also been the subject of strong debate 
\cite{shevchenko,hyodo,sato,akaishi:2007cs}.

    Experimentally, the great excitement generated by peaks seen at KEK  
\cite{Suzuki:2004ep}  and FINUDA \cite{Agnello:2005qj,:2007ph}, originally
interpreted in terms of deeply bound kaons atoms, has calmed down, particularly
after the work of \cite{toki} regarding the KEK experiment and of 
 \cite{Magas:2006fn,Crimea,Magas:2008bp} regarding the FINUDA ones found 
 explanations of the  experimental
 peaks based on conventional reactions that unavoidably occur in the process of
 kaon absorption. Also the thoughts of \cite{Suzuki:2007kn}, with opposite views
 to those of FINUDA in \cite{Agnello:2005qj}, and the reanalysis of the KEK
 experiment of \cite{Suzuki:2004ep} done in \cite{Suzuki:2007kn}, where the
 original narrow peak appears  much broader, have helped to bring the discussion
 to more realistic terms. In any case, this short discussion has served to show
 the intense activity and interest in the subject.
 
     There is however one experiment from where the authors claim  
evidence for a very strong kaon-nucleons potential, of the order of 200 MeV
attraction \cite{Kishimoto:2007zz}.  The experiment looks for fast protons
coming out from the absorption of kaons in flight in nuclei. The problem has
been recently studied in \cite{magasflight} where it was 
shown how the experiment was analyzed and which ingredients are missing in the
analysis of \cite{Kishimoto:2007zz}. 
   
   One of the shortcomings of \cite{Kishimoto:2007zz} stems from the use of the
Green's function method \cite{Morimatsu:1994sx} in a variant used in 
\cite{YH,Yamagata:2007cp,YamagataSekihara:2008ji} to analyze the data and extract
from there the kaon optical potential.  The only mechanism considered in 
\cite{Kishimoto:2007zz} for the emission of fast protons is the $\bar{K} N \to   
\bar{K} N$, taking into account the optical potential for the kaon in the final 
state. However, in \cite{magasflight} one can see that there are other 
mechanisms that
contribute to the emission of fast protons.  One of the mechanisms is the kaon
absorption by one nucleon, $K^- N \to \pi \Sigma$ or $K^- N \to \pi \Lambda$
followed by decay of the $\Sigma$ or the $\Lambda$ into $\pi N$. Another of 
the mechanisms is the
absorption  of kaons by pairs of nucleons, $\bar{K} N N\to  \Sigma N$
and $\bar{K} N N\to  \Lambda N$, followed by similar hyperon decays. 
The contributions from these processes were also suggested in Ref. \cite{YH}.
  Another important point disclosed in  \cite{magasflight} is that in the
experiment of \cite{Kishimoto:2007zz} there was an extra requirement of
coincidence: in addition to a fast proton detected forward one demanded that
there would be at least a charged particle detected in a layer counter 
surrounding the target.  It was assumed in \cite{Kishimoto:2007zz} that this
does not change the shape of the proton distribution, but it was found in 
\cite{magasflight} that this is not the case and the shape of the spectrum changes substantially, to the point of invalidating the conclusions drawn
in \cite{Kishimoto:2007zz}. Although in \cite{magasflight} one does not make
claims for a certain strength of the kaon potential, because the reaction is
not particularly suited to determine this magnitude, one at least finds that a
shallow potential is preferred to a very strong one. 

\section{Correlated $\Lambda d$ pairs emitted after the absorption of kaons at
rest in nuclei}

As an example of how one does interpret experiments which have been advocated as
evidence for deeply bound kaon atoms in nuclei we explain here the 
case in favor of a deeply bound kaon from a peak seen in the experiment of 
\cite{:2007ph}. In this experiment $\Lambda$ and $d$ were measured in
coincidence after kaon absorption at rest from $^6$Li and $^{12}$C, and it was
observed that the $\Lambda$ and $d$ were strongly correlated back to back. From
there the authors concluded the formation of a bound state of a kaon and three
nucleons which decays in $\Lambda~d$. 

In the $K^{-}_{stop} A \to \Lambda d A'$ reaction \cite{:2007ph},  at least
three nucleons must participate in the absorption process.  Two body $K^-$
absorption processes of the type $K^- NN \to \Sigma N ( \Lambda N) $  have been
studied experimentally in \cite{Katz:1970ng} and their strength is seen to be
smaller than that of the one body absorption  $K^- N \to \pi \Sigma (\pi
\Lambda )$ mechanisms. This result follows the  argument that it is easier to
find one nucleon than two nucleons together in the nucleus. This is also the
case in pion absorption in nuclei, where extensive studies, both theoretical
\cite{Oset:1986yi} and experimental \cite{Weyer:1990ye},  obtain the direct two
and three body absorption rates with the former one dominating over the later,
particularly for pions of low  energy. We follow here the same logics and
assume the process to be dominated by direct three body $K^-$ absorption, the
four body playing a minor role. 

   The former assumption means in practice that, in $^6$Li, the other three nucleons not
directly involved in the absorption process will be spectators in the reaction.
  These three spectator nucleons have to leave the nucleus, but
  they were originally bound in the nucleus. The nuclear dynamics takes care of this since 
  there is a distribution of momenta and energies in the nucleus, and the
  ejection of either three nucleons, a $n d$  pair or tritium, implies that the
  absorption is done in the most bound nucleons. 
  
  The other element of relevance
  is the atomic orbit from which the kaon is absorbed. This information is 
  provided
  by the last measured transition in the X-ray spectroscopy of $K^-$-atoms, 
  which occurs precisely
  because absorption overcomes the $\gamma$ ray emission. In the case of $^6$Li
  this happens for the $2p$
  atomic state \cite{okumura,friedman-gal}.  
  
  Following the line of studies done for pion absorption 
  and other inclusive reactions \cite{Salcedo:1987md}, 
 we describe the nucleus in terms of a local Fermi sea with Fermi momentum
 $k_F(r)$. The nucleons move in a mean field
given by the Thomas Fermi potential
 \beq
V(r)=-\frac{k_F^2(r)}{2m_N}\,, \quad k_F(r)=\left(\frac{3\pi^2}{2}\rho(r) \right)^{1/3}\,,
\eeq{e1}
 where $m_N$ is the nucleon mass and $\rho(r)$ is the local nucleon density 
 inside the nucleus. 
 
 This potential assumes a continuity from the energies
 of the bound states (holes) to those in  the continuum (particles), which is
 not the case in real nuclei. For this reason, we implement an energy gap, 
 $\Delta$, which is adjusted to respect the threshold of
 the reaction. The introduction of a gap in the Fermi sea is a common practice 
 in order to be precise with the actual binding
 energies of the  nuclei involved in a particular reaction so that the corresponding threshold is respected 
 \cite{Kosmas:1996fh,Albertus:2001pb,Nieves:2004wx}.
 Hence, we demand that the highest possible invariant mass of 
  $K^- NNN$ system, which happens when the three nucleons are at the
 Fermi surface with total three-momentum zero, corresponds to the minimum
 possible energy for a spectator three-nucleon system with total zero momentum,
 namely a tritium at rest. 
   This situation corresponds to
  \beq
  m_{K^-} + M_{ ^6{\rm Li}}= 
  m_{K^-}+3m_N-3\Delta+M_t \ ,
  \eeq{e2}
 and we determine $\Delta=7.8$ MeV. In the above expression $m_{K^-}$,  $M_t$,
 $M_{ ^6{\rm Li}}$ are the masses of the 
 corresponding particles and nuclei. 
   
   The probability of $K^-$ absorption by three nucleons will be determined from
   the third power of the nuclear density as
  \beq
 \Gamma \propto \int d^3 \vec{r}  |\Psi_{K^-}(r)|^2 \rho^3(r)\,,
  \eeq{e3}
  where $\Psi_{K^-}(r)$ is the $K^-$ atomic wave function.
   In order to take into account the Fermi motion we write the density as 
$\rho(r)=4 \int \frac{d^3 \vec{p}}{(2\pi)^3} \Theta(k_F(r)-|\vec{p}|)$ 
   and then we obtain
$$ 
\Gamma \propto \int d^3 \vec{r} d^3 \vec{p}_1  d^3 \vec{p}_2 d^3 \vec{p}_3 |\Psi_{K^-}(r)|^2\times
$$
\beq
 \times \Theta(k_F(r)-|\vec{p}_1|)\Theta(k_F(r)-|\vec{p}_2|)\Theta(k_F(r)-|\vec{p}_3|)\,.  
\eeq{P}

   From this expression we can evaluate all observables of the reaction. Let us first
   concentrate on the $\Lambda d$ invariant mass which, for each $K^- NNN \to
   \Lambda d$ decay event, is precisely the invariant mass of the 
   corresponding $K^-NNN$ system, the other three nucleons acting as spectators.
   Thus the energy of the  $\Lambda d$ pair is obtained from
   \begin{eqnarray}
 &&E_{\Lambda d} = E_{K^-NNN} \equiv E_{K^-}+ E_{N_1} + E_{N_2} + E_{N_3}\\
   &&\!\!\!\! =
   \!m_{K^-}\!+\!3m_N\! +\! \frac{\vec{p}_1^2}{2m_N}\!+\!
   \frac{\vec{p}_2^2}{2m_N}\!+\!
   \frac{\vec{p}_3^2}{2m_N}\!-\!3\frac{k_F^2(r)}{2m_N}\!-\!3\Delta \ , \nonumber 
   \label{e6a}
 \end{eqnarray}
 and the momentum from
   \beq
   \vec{P}_{\Lambda d}=\vec{P}_{K^-NNN} =
   \vec{p}_1+\vec{p}_2+\vec{p}_3 \ ,
   \eeq{e6b}
and, correspondingly,
   \beq
   M_{\Lambda d} = E_{\Lambda d} - \frac{\vec{P}_{\Lambda d}^2}{2E_{\Lambda d}}\ .
   \eeq{e6}
   One may also easily obtain the invariant mass of the residual system,
   $M^*$, from
  \begin{equation}
     M^* = E^* - \frac{\vec{P}^{*\,2}}{2E^*}\ ,
     \label{mst}
  \end{equation}
  with
   \beq
   E^*=m_{K^-}+M_{ ^6{\rm Li}}-E_{K^-NNN}\ , \quad \vec{P}^*=-\vec{P}_{\Lambda d}\ .
    \eeq{aa}

  Each event in the multiple integral of Eq.~(\ref{P}), done with the Monte Carlo
  method, selects particular values for $\vec{r}$, $\vec{p}_1$, $\vec{p}_2$ and 
  $\vec{p}_3$ which, in turn, determine the value of the corresponding $\Lambda d$ 
  invariant mass from Eqs.~(\ref{e6a})--(\ref{e6}). 
   Since the minimum obvious invariant mass of the residual
     three-nucleon system is $M^*=M_t$, corresponding to the emission of
     tritium, the cut $\Theta(M^* - M_t)$ is also imposed for each event. 
  A compilation of events provides us with
   the $\Lambda d$ invariant mass distribution.    
   We also directly obtain the distribution of
   total $\Lambda d$ momentum, Eq.~(\ref{e6b}), to be directly compared with 
   the $\Lambda d$ momentum measured in \cite{:2007ph}. 
   
   Note that the model presented here is a straightforward generalization (from two nucleon to three nucleon $K^-$ absorption)
    of the one used in Refs. \cite{Magas:2006fn,Ramos:2007zz}, however here we concentrate on the primary reaction peak, while in Refs. \cite{Magas:2006fn,Ramos:2007zz}     the authors were more interested in the peak generated by the final state interactions, i.e. by the collisions of the primary produced $\Lambda$     and $p$ on their way out of the nucleus. Since the two nucleon $K^-$ absorption, discussed in \cite{Magas:2006fn,Ramos:2007zz}, was measured for heavier    nuclei \cite{Agnello:2005qj}, the final state interaction peak was stronger than that of the primary reaction, contrary to the reaction studied in       this work. 
    
   Other observables measured in \cite{:2007ph} require an additional work.
   One is the angular
   correlation of $\Lambda d$ pairs, and the other is the missing mass 
   assuming a residual $n d$ system, apart from the measured $\Lambda d$ pair, 
   namely
   \beq
   T_{miss} = m_{K^-}+M_{ ^{6}{\rm Li}} - m_\Lambda - m_n - 2 M_d - (T_\Lambda + T_d) 
   \ ,
   \eeq{e7}
   where $m_\Lambda$, $M_d$ and $T_{\Lambda}$, $T_d$ are the masses and the 
   kinetic energies of the  $\Lambda$ and the $d$, correspondingly. These two
   observables require the evaluation of the individual $\Lambda$ and $d$
   momenta in the laboratory frame. Their value in 
   the center of mass (CM) frame of the $\Lambda d$ pair is given
   in terms of the known invariant mass but their direction in this frame is
   arbitrary. We take this into account by obtaining $\Lambda$ and $d$ momenta
   in the CM frame
     \begin{eqnarray}
  \vec{p}_\Lambda^{CM}&=&p_{\Lambda}^{CM}\left(\sin\Theta \cos\phi, 
  \sin\Theta \sin\phi, \cos\Theta \right)\ , \nonumber \\
   \vec{p}_d^{CM}&=&-\vec{p}_\Lambda^{CM} \ ,
  \end{eqnarray}
  with
    \beq
  p_\Lambda^{CM}=\frac{\lambda^{1/2}(M_{\Lambda
  d}^2,m_{\Lambda}^2,M_d^2)}{2M_{\Lambda d}}\ , 
  \eeq{e8}
  where the events are now generated according to the distribution provided by
  the integral 
\begin{eqnarray}
&& \int d \cos\Theta \int d\phi \int d^3 \vec{r} d^3 \vec{p}_1  d^3 \vec{p}_2 d^3 \vec{p}_3 
 |\Psi_{K^-}(r)|^2 \nonumber \\ 
&& \times \Theta(k_F(r)-|\vec{p}_1|)
 \Theta(k_F(r)-|\vec{p}_2|)\Theta(k_F(r)-|\vec{p}_3|)\nonumber \\
 && \times \Theta(M^* - M_t) \ .
  \end{eqnarray}
In order to have the final $\Lambda$ and $d$ momenta in the laboratory frame,
where the $\Lambda d$ pair has momentum $\vec{P}_{\Lambda d}$, we apply the 
transformations
  \begin{eqnarray}
  \vec{p}_\Lambda &=&\vec{p}_\Lambda^{CM}+ m_\Lambda \vec{v} \nonumber \\
 \vec{p}_d&=& -\vec{p}_\Lambda^{CM}+ M_d \vec{v}\ ,
  \end{eqnarray}
  where $\vec{v}=\vec{P}_{\Lambda d}/(m_\Lambda+M_d)$.
  These last equations allow us to find the cosinus of the angle between 
  the directions of $\Lambda $ and $d$. Therefore, generating the 
  distribution of events according to their relative angle is
  straightforward. We see, as it is also the case of the experiment,
  that  $P_{\Lambda d}\sim 200$ MeV/c, while $p_{\Lambda}^{CM} \sim 650$ 
  MeV/c, which already guarantees that the $\Lambda d$ events are largely 
  correlated back-to-back.
   
    We note that our calculations incorporate the same momentum cuts as
    in the experiment, namely  140  MeV/c $< p_{\Lambda}<$ 700 MeV/c
 and  300  MeV/c $< p_d <$ 800  MeV/c.

\begin{figure*}[!htb]
\vspace{-0.0cm}
\centering
\includegraphics[width=8.4cm]{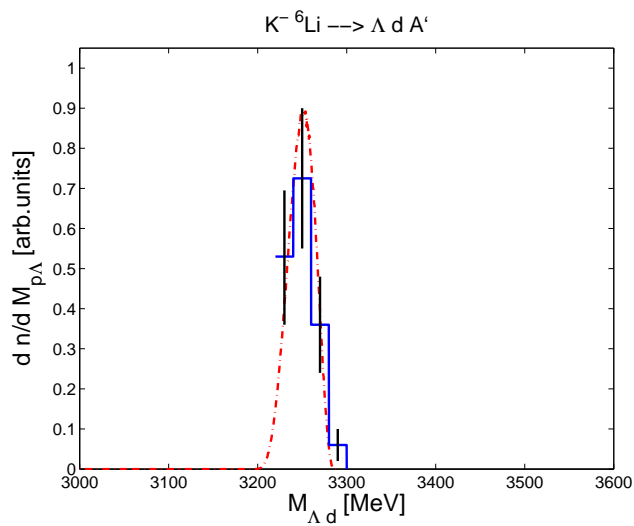}
\vspace{-0.0cm}
 \caption{(Color online) The $\Lambda d$ invariant mass distribution for the $K^{-}_{stop} A
\to \Lambda d A'$ reaction. Histogram and error bars are from the experimental
paper \cite{:2007ph}, while the dot-dashed curve is the result of our calculation.}
\label{fig1}
\vspace{-0.0cm}
\end{figure*}

 In Fig. \ref{fig1} we show the results for the invariant mass of the $\Lambda d$ system.
 Our distribution, displayed with a dot-dashed line, peaks around 
 $M_{\Lambda d}= 3252$ MeV as in the experiment. The shape of the distribution
 also compares remarkably well with the experimental histogram in the region of
 the peak, which is the energy range that we are exploring in the present work. 
 We obtain a width of about $36$ MeV, as reported in the
 experiment. Note that apart from the peak that we are discussing, the
 experiment also finds events at lower $\Lambda d$ invariant masses which did
 not play a role in their discussion \cite{:2007ph}. These
 events would be generated in cases where there is final state interaction of
 the $\Lambda$ or the $d$  with the
 rest of the nucleons, as was discussed in \cite{Magas:2006fn,Ramos:2007zz}, or through 
 other absorption mechanisms. 
 
\begin{figure*}[!htb]
\vspace{-0.0cm}
\centering
\includegraphics[width=8.4cm]{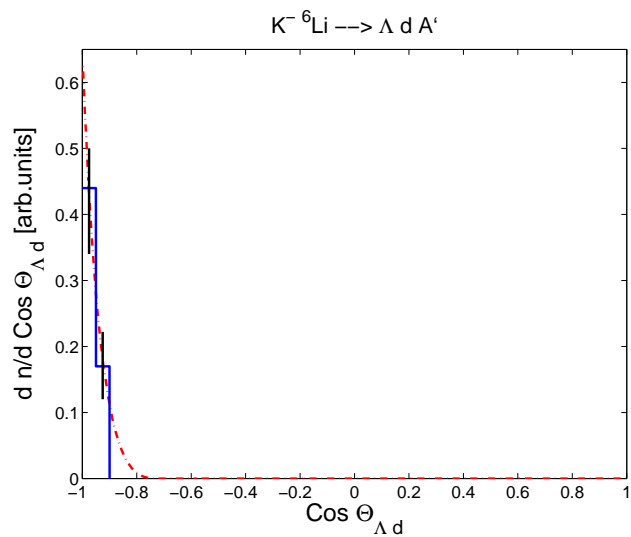}
\vspace{-0.0cm}
\caption{(Color online) The $\Lambda d$ angular distribution for the $K^{-}_{stop} A \to
\Lambda d A'$ reaction. Histogram and error bars are from the experiment
\cite{:2007ph}, while the dot-dashed curve is the result of our calculation. 
As in the experimental analysis, we take into account the following cuts: 
 3220 MeV $ < M_{\Lambda d} <$ 3280  MeV.}
\label{fig2}
\vspace{-0.0cm}
\end{figure*}

  The angular correlations between the emitted $\Lambda$ and $d$ can be seen in
  the distribution displayed in Fig. \ref{fig2}, where, as
  in the experimental analysis, we consider only those events
  which fall  in the region  3220  MeV $< M_{\Lambda d} <$ 3280  MeV.
  As we can see in the figure, the
  distribution is strongly peaked backward and the agreement with experiment is
  very good.

    One also finds good agreement with data for the distribution of the total 
    $\Lambda d$ momentum and the missing mass distribution.

\section{The $\bar{K} NN$ system}

In the context of low-energy QCD with $N_f = 3$ quark flavors, the study of 
possible antikaon-nuclear quasibound states is a topic of great current 
interest. Spontaneously broken chiral $SU(3)\times SU(3)$ symmetry, together
with explicit symmetry breaking by the non-zero quark masses, basically 
determines the leading couplings between the low-mass pseudoscalar meson 
octet (Nambu-Goldstone bosons in the chiral limit) and the octet of the 
ground state baryons. In particular, the Tomozawa-Weinberg chiral low-energy
theorem implies that the driving $\bar{K}N$ interaction in the isospin $I=0$
channel is  strongly attractive. Likewise, the $I=0$ $\pi\Sigma$ interaction
is attractive. The coupling between these $\bar{K}N$ and $\pi\Sigma$ 
channels is the prime feature governing the subthreshold extrapolation of 
the $\bar{K}N$ interaction.

We have already discussed the problem of the $K^-$-nucleus interaction and $K^-$ bound states. A prototype system for these considerations is $ppK^-$, the 
simplest antikaon-nuclear cluster. It has recently been investigated using 
three-body methods with Faddeev equations \cite{shevchenko,sato} and 
variational approaches 
 \cite{akaishi:2007cs,DW-HYP06,ppK_Lstar:Oka,DHW:2008}.
Reaction studies~\cite{ppK_Reac:Koike07} have also been performed dealing 
with the actual formation of $ppK^-$. The Faddeev and variational 
calculations predict a total $ppK^-$ binding energy in a range $B\sim$ 
50 - 70 MeV, together with an estimate of the $\bar{K}NN\rightarrow\pi Y N$ 
decay width, $\Gamma \sim$ 50~-~100 MeV, depending on details of the 
interactions used.

The key issue in any such calculation is the extrapolation
of the $\bar{K}N$ interaction into the region far below threshold. Its 
predictive power is limited by the lack of accurate 
constraints from data. 

Apart from the constraints provided by $\bar{K}N$ threshold data and 
low-energy cross sections, the only piece of information about the 
interaction below $\bar{K}N$ threshold is the $\pi\Sigma$ mass spectrum 
which is dominated by the $\Lambda(1405)$ resonance. In \cite{hyodo} the 
extrapolation below threshold of the effective $\bar{K}N$ interaction has been 
done following  ~\cite{Hyodo:2007jq} from the viewpoint of chiral 
SU(3) dynamics. Their structure shares features with the 
pioneering coupled-channel 
model ~\cite{Dalitz:1967fp} that used vector meson exchange interactions
(see also Ref.~\cite{Siegel:1988rq}).

Most chiral SU(3) based calculations produce two $\Lambda(1405)$ states which are seen in the  $\pi\Sigma$ mass spectrum \cite{Jido:2003cb}. The higher mass state peaks around 1420 MeV and is relatively narrow, while the lower mass state is more uncertain but peaks around 1395 MeV and is much wider. This implies that the effective 
single-channel $\bar{K}N$ interaction is substantially weaker than 
anticipated in the simple phenomenological potential of \cite{akaishi:2002bg,akainew}. In those phenomenological studies, the 
local, energy-independent potential was adjusted interpreting the 
$\Lambda(1405)$ directly as a $\bar{K}N$ bound state, identifying its 
binding energy by the location of the maximum observed in the $\pi\Sigma$ 
spectrum of the reaction of \cite{Thomas:1973uh}, but ignoring strong coupled-channel effects.

In \cite{hyodo} a variational $ppK^-$ calculation was performed employing the new
effective $\bar{K}N$ potential derived from chiral coupled-channel 
dynamics in ~\cite{Hyodo:2007jq}, together with a realistic $NN$ potential. This
calculation is supposed to be complementary to the Faddeev approach with 
chiral SU(3) constraints~\cite{sato}. The variational calculation 
gives easy access to the wave function of the bound state so that valuable 
information about the structure of the $ppK^-$ 
cluster can be extracted, whereas the elimination of the $\pi\Sigma$ channel
is required and the width of the state can only be estimated perturbatively.
The Faddeev calculation has, in turn, the advantage that the decay width of 
the quasibound state is computed consistently in the coupled-channel 
framework. Both methods therefore have their virtues and limitations.

The work of \cite{hyodo}
 extends and improves the previous studies~\cite{DW-HYP06} 
in several directions, including further refinements in the $NN$ 
interaction, computation of density distributions, an evaluation of 
effects from $p$-wave $\bar{K}N$ interactions and an estimate of the 
$\bar{K}NN \rightarrow YN$ absorptive width.

  The present situation is rather uncertain and the values of the bindings vary from about 20 to 70 MeV and the width from about 50 to 100 MeV. The different way in which the two body amplitudes are extrapolated off shell might be partly responsible in these three body calculations for the differences in those approaches.  In this sense, in section \ref{sec:1} we shall present a method to deal with Faddeev equations in coupled channels which eliminates from the beginning this unphysical part of the scattering amplitudes.

\section{Conclusion on experimental situation on deeply bound kaon clusters}
No evidence of deeply bound $K^-$ states on nuclei has been found.
All peaks claimed as states could be interpreted in terms of conventional,
unavoidable and controllable reactions.
Work continues searching for these states in
JPARC, FINUDA, AMADEUS, DISTO 
 (see talks by 
 Shevchenko,  Morton, Grishina, Vazquez Poce, Lio, Okada
Camerini and Tsukada in this Conference).
On the positive side, we are learning new and interesting physics
about $\bar{K}$ absorption by two and three nucleons, which is relevant from the many body point of view. 

\section{Other kaonic clusters}
  In this section I want to call the attention to new and very interesting
kaonic clusters which have been investigated only in the last couple of years. I
will mention the system with two meson and a baryon, with one of the mesons being
a kaon. Some of these three body systems are bound, or form resonances, and lead
to the low lying $1/2^+$ excited baryons with strangeness.  The other systems
are also three body systems with a cluster of  $K \bar{K}$, one of them with an
extra nucleon, another one with an extra $\phi$ meson and the third one with
a $J/\psi$.  The study of these new systems has been stimulated by a
reformulation of the Faddeev equations within the formalism of the chiral
unitary approach \cite{mko,MartinezTorres:2008is}.

\subsection{Formalism}
\label{sec:1}
We solve Faddeev equations in the coupled channel approach. These coupled systems have first been constructed by pairing up all the possible pseudoscalar mesons and $1/2^+$ baryons, which couple to strangeness =-1, and adding a pion to the pair finally. One ends up with twenty-two coupled channels for a fixed total charge \cite{mko}.
The interaction in the  meson-baryon sub-systems has been written in  $S$-wave, which implies, after adding another meson in $S$-wave, that the total $J^P$ of the three body system is $1/2^+$. 

The solution of the Faddeev equations,
\begin{equation}\label{ft}
T = T^1 + T^2 + T^3,
\end{equation}
where the $T^i$ partitions are written in terms of two body t-matrices ($t^i$) and three body propagators
($g = [E - H]^{-1}$) as
\begin{equation}\label{fp}
T^{i}=t^i + t^i g \Big[ T^{j} + T^{k} \Big] \,\,\,\, {\rm (i \neq j, \,\,i,j \neq k =1,2,3)},
\end{equation}
requires off-shell two body $t$-matrices as an input. However, we found that 
the off-shell contributions of the two body $t$-matrices give rise to 
three-body forces which, when summed up for different diagrams, cancel the one 
arising directly from the chiral Lagrangian in the $SU(3)$ limit \cite{mko}. In the realistic case, 
this sum was found to be only 5 \% of the total on-shell contribution \cite{mko}. Thus, 
it is reasonably accurate to study the problem by solving the Faddeev equations with the on-shell two-body $t$-matrices and by neglecting the three-body forces.

\begin{figure*}[!htb]
\centering
\resizebox{0.25\textwidth}{!}{%
\includegraphics{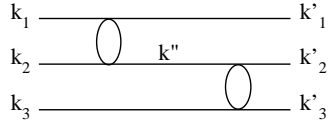}
}
\caption{\textit{The diagrammatic representation of the $t^1gt^3$ term.}}
\label{figtgt}
\end{figure*}

In this way, a term at second order in $t$ in the Faddeev partitions 
(fig. \ref{figtgt}), for instance, 
\begin{equation}\label{t1gt3}
t^1 g^{13} t^3,
\end{equation}
is written as a product of the on-shell two body $t$-matrices $t^1$ and $t^3$
and the propagator $g^{13}$ given as
\begin{eqnarray}
g^{13} = \frac{1}{2E_2}\frac{1}{\sqrt{s}-E_1
(\vec{k}^\prime_1)-E_2(\vec{k}^\prime_1+\vec{k}_3)-E_3(\vec{k}_3)+i\epsilon}\nonumber.
\end{eqnarray}

\begin{figure*}[!htb]
\centering
\resizebox{0.25\textwidth}{!}{%
\includegraphics{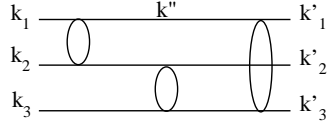}
}
\caption{\textit{The diagrammatic representation of $t^2g^{21}t^1g^{13}t^3$.}}
\label{tgtgt}
\end{figure*}

Adding another interaction to the diagram in fig. \ref{figtgt} (see fig. \ref{tgtgt}),
 the expression $t^1g^{13}t^3$
gets extended to\\
$t^2g^{21}t^1g^{13}t^3$, which can be written explicitly as
\begin{eqnarray}
&&t^2(s_{13})\Bigg[\int \frac{d\vec{k}^ {\prime\prime}}{(2\pi)^3} \frac{1}{2E_1(\vec{k}^{\prime\prime})} \frac{1}{2E_2(\vec{k}^{\prime\prime}+\vec{k}_3)} \frac{2 M_3}{2 E_3(\vec{k}^{\prime\prime}+ \vec{k}^\prime_2)}\nonumber\\
&&\times\frac{t^1(s_{23}(\vec{k}^{\prime\prime}))}{\sqrt{s}-E_1
(\vec{k}^{\prime\prime})-E_2(\vec{k}^{\prime}_2)-E_3(\vec{k}^{\prime\prime}+\vec{k}^{\prime}_2)+i\epsilon}\nonumber\\
&&\times\frac{1}{\sqrt{s}-E_1(\vec{k}^{\prime\prime})-E_2(\vec{k}^{\prime\prime}+\vec{k}_3)-E_3(\vec{k}_3)+i\epsilon}\Bigg]t^3(s_{12}),\nonumber
\end{eqnarray}
where $s_{13} = (P - K^{\prime}_2)^2$, $s_{12} = (P - K_3)^2$, $s_{23}(\vec{k}^{\prime\prime}) = (P - K^{\prime\prime})^2$ with $P, K_3, K^{\prime}_2$ and $K^{\prime\prime}$ representing the four momenta corresponding to the $\vec{P} = 0, \vec{k}_3, \vec{k}^{\prime}_2$ and $\vec{k^{\prime\prime}}$ respectively.
Our aim is to extract $t^1g^{13}t^3$ out of the integral, which could simplify the calculations. The $t^2(s_{13})$ and $t^3(s_{12})$, in the equation above , depend on on-shell variables and can be factorized out of the loop integral but not $t^1$ and $g^{13}$ \cite{mko} . This can be done if we re-arrange the loop integral as is done in \cite{mko}.

There one defines  a function  $F^{ijk}$ which includes the loop variable dependence of $t^j$. Then one defines a function $G^{ijk}$ as

\begin{eqnarray}
G^{213} = \int \frac{d\vec{k}^{\prime\prime}}{(2\pi)^3}\frac{1}{2E_1(\vec{k}^{\prime\prime})}\frac{2M_3}{2E_3(\vec{k}^{\prime\prime})}\frac{F^{213}(\sqrt{s},\vec{k}^{\prime\prime})}{\sqrt{s_{13}} - E_1(\vec{k}^{\prime\prime}) - E_3(\vec{k}^{\prime\prime})+i\epsilon}\nonumber
\end{eqnarray}
The diagram in fig. \ref{tgtgt} can, hence, be re-written as $t^2G^{213}$ 
$t^1g^{13}t^3$.
The formalism has been developed following the above procedure, i.e., by 
replacing $g$ by $G$, every time a new interaction is added. This leads to 
another form of the Faddeev partitions (eq. (\ref{fp})), which, after removing 
the terms corresponding to the disconnected diagrams and by denoting the rest 
of the equation as $T_R$, can be re-written as 
\begin{equation}
T_R^{ij}=t^i g^{ij} t^j + t^i G^{ijk} T_R^{jk} + t^i G^{iji} T_R^{ji}, \,\,\,\, {\rm i \neq j, j \neq k =1,2,3}.
\end{equation}

The $T_R^{ij}$ can be related to the Faddeev partitions (eq. (\ref{fp})) 
as $T^i = t^i + T_R^{ij} + T_R^{ik}$, hence, giving six coupled equations 
instead of three (eq. (\ref{fp})). These $T_R^{ij}$ partitions correspond to 
the sum of all the possible diagrams with the last two interactions written 
in terms of $t^i$ and $t^j$. This re-grouping of diagrams is done for the sake 
of convenience due to the different forms of the $G^{ijk}$ functions. We 
define $T_R$ as 
\begin{equation}\label{tr}
T_R = \sum_{i \neq j = 1}^3 T_R^{ij},
\end{equation}
which can be related to the sum of the Faddeev partitions (eq. (\ref{ft})) as 
\begin{equation}
T = t^1 + t^2 + t^3 + T_R.
\end{equation}

\subsection{Results and discussion}\label{sec:2}
One constructs the three-body $T_R$-matrices using the isospin symmetry, 
for which we must take an average mass for the isospin multiplets 
$\pi$ $(\pi^+,\,\pi^0,\,\pi^-)$, $\bar{K}$ $(\bar{K}^0,\,K^-)$, $K$ 
$(K^+,\, K^0)$, $N$ $(p,\,n)$, $\Sigma$ $(\Sigma^+,\,\Sigma^0,\,\Sigma^-)$ and 
$\Xi$ $(\Xi^0,\,\Xi^-)$. In order to identify the nature of the resulting states,
 we project the $T_R$-matrix on the isospin base. One appropriate base is the 
 one where the states are classified by the total isospin of the three 
 particles, $I$, and the total isospin of the two mesons, $I_\pi$ in the case 
 of two pions. Using the phase convention $\mid \pi^+\rangle=-\mid 1,1\rangle$, 
 $\mid K^-\rangle=-\mid 1/2,-1/2\rangle$, $\mid \Sigma^+\rangle=-\mid 1,1\rangle$ 
 and $\mid \Xi^-\rangle=-\mid 1/2,-1/2\rangle$ we have, for example, for the 
 the $\pi\,\pi\,\Sigma$ channel
\begin{eqnarray}
\nonumber
\mid \pi^0\,&\pi^0&\,\Sigma^0\rangle =\mid 1, 0\rangle_{\pi}\,\otimes\mid 1, 0\rangle_{\pi}\,\otimes\mid 1, 0\rangle_{\Sigma} \nonumber\\
&=&\left\{\sqrt{\frac{2}{3}}\mid 2, 0\rangle-\sqrt{\frac{1}{3}}\mid 0, 0\rangle\right\}_{\pi\pi}\,\otimes
 \mid 1, 0\rangle_{\Sigma}\nonumber\\
&=&\sqrt{\frac{2}{5}}\mid I=3,I_\pi=2\rangle-\frac{2}{\sqrt{15}}\mid I=1,I_\pi=2\rangle-\nonumber\\
&&-\sqrt{\frac{1}{3}}\mid I=1,I_\pi=0\rangle,\nonumber
\end{eqnarray}
where, $I$ and $I_{\pi}$ denote the total isospin of the three body system and 
that of the two pion system, respectively. Similarly, we write the other 
states in the isospin base. 

\begin{figure*}[!htb]
\resizebox{1\textwidth}{!}{%
  \includegraphics{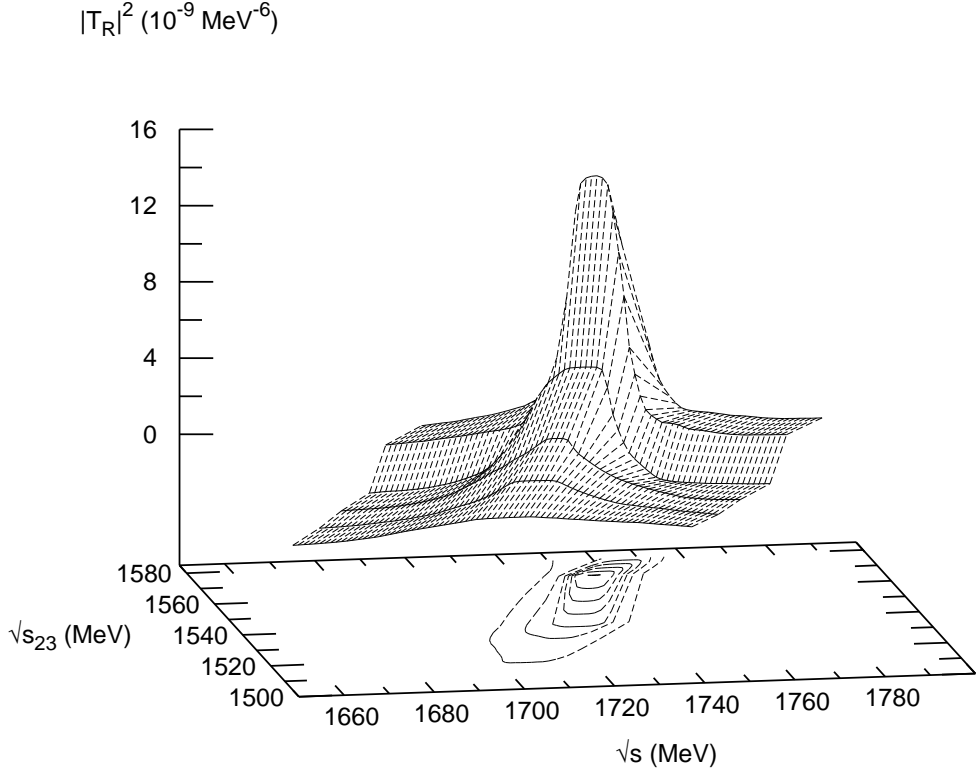}
}
\caption{The $\Lambda$(1810) resonance depicted in the squared amplitude for 
$\pi \pi \Lambda$ in the $\mid I, I_{\pi}\rangle$ = $\mid 0, 0\rangle$ configuration.}
\label{fig:1}       
\end{figure*}

After projecting the $T_R$-matrix (eq. (\ref{tr})) on the isospin base, we 
square them and plot them as a function of the total energy ($\sqrt{s}$) of 
the two meson one baryon system and the invariant mass of particles $2$ and 
$3$ ($\sqrt{s_{23}}$). However, this choice is arbitrary, we could as well plot 
the squared amplitude, for example,  as a function of $\sqrt{s_{12}}$ and 
$\sqrt{s_{13}}$.

In fig. \ref{fig:1}, we show our results for $\pi \pi \Lambda$ in the 
$\mid I, I_{\pi}\rangle$ = $\mid 0, 0\rangle$ case, where a peak at 1740 MeV 
is clearly seen. The full width at half maximum of this peak is $\sim$ 20 MeV. 
We identify this resonant structure with the $\Lambda$(1810) in the particle 
data book \cite{pdg}.  It should be noted that, even though the status of this 
resonance in \cite{pdg} is three-star, the associated pole positions vary from 
about 1750 MeV to 1850 MeV and the corresponding widths from 50-250 MeV. 

We find evidence for two more resonances in the isospin zero sector in the 
$\Lambda$(1600) region. Apart from this, we find evidence for four low-lying 
$\Sigma$ resonances: 1) for the $\Sigma$(1560), thus predicting $J^P =1/2^+$ 
for it, 2) the $\Sigma$(1620), for which the partial-wave analyses and 
experimental findings are kept separately in the $PDG$, 3) the well-established 
$\Sigma$(1660) and 4) the one-star $\Sigma$(1770). All these results are 
summarized in the table 1, along with the states listed by the Particle Data 
Group.

\subsection{The $\bar{K} \pi N $ and coupled channels system}
  In \cite{mko} a total of 22 coupled channels to $\bar{K} \pi N $ were
  considered. The T matrix was projected over total isospin states  and  
  the $|T|^2$ matrix was plotted against two variables, the total energy and the
  invariant mass of a pair of particles, either a meson and a baryon of the two
  mesons. The magnitude $|T|^2$ exhibits clear peaks around a value of $\sqrt
  s_{23}$ and $\sqrt s$. This tell us not only that there is a resonance at a
  certain energy but also that the pair of particles considered are highly
  correlated around th energy at the peak. Usually this corresponds to the energy
  where the pair considered forms a dynamically generated resonance from meson
  baryon like the $\Lambda(1405)$ or from two mesons ($K \bar{K}$ and $\pi \pi$)
  like the $f_0(980)$. In table \ref{tab:1} we show the summary of the results
  obtained

  \begin{table}[h]
\begin{center}
\caption{A comparison of the resonances found in this work with the states in $PDG$.}
\label{tab:1}       
\vspace{0.5 cm}
\begin{tabular}{lccc}
\hline\noalign{\smallskip}
& $\Gamma$ ($PDG$) & Peak position  & $\Gamma$ (this work)\\
& (MeV) & (this work, MeV) & (MeV)\\
\noalign{\smallskip}\hline\noalign{\smallskip}
\multicolumn{4}{l}{Isospin = 1} \\
\noalign{\smallskip}\hline\noalign{\smallskip}
$\Sigma(1560)$&10 - 100&1590&70\\
$\Sigma(1620)$&10 - 100&1630&39\\
$\Sigma(1660)$&40 - 200&1656&30\\
$\Sigma(1770)$&60 - 100&1790&24\\
\noalign{\smallskip}\hline
\multicolumn{4}{l}{Isospin = 0} \\
\noalign{\smallskip}\hline
$\Lambda(1600)$&50 - 250&1568,1700 &60,135\\
$\Lambda(1810)$&50 - 250&1740&20\\
\noalign{\smallskip}\hline

\end{tabular}
 \end{center}
\end{table}

\subsection{State of $K \bar{K} N$ and coupled channels}

In \cite{jidoenyo} using a variational method and chiral dynamics for the
interaction of the kaons with nucleons a bound state of the $K \bar{K} N$ system
was found, where the $K \bar{K}$ system was forming the $a_0(980)$ resonance.
  The importance of using coupled channels has been made patent in
\cite{mko,MartinezTorres:2008is}.  In this sense the idea of \cite{jidoenyo} was
retaken in \cite{albertopheno} using coupled channels.  The paper also had
a novel idea. The fact that it was found that there was a cancellation
between the off shell part of the two body amplitudes and the three body
amplitudes coming from the same chiral Lagrangians was understood as some
thing more profound, that in this problems one can use on shell amplitudes
and not worry about three body forces, and of course the on shell two
body amplitudes can be obtained from experiment and used beyond the
region of applicability of the chiral unitary approach that relies only
on the lowest order chiral Lagrangian and absence of possible genuine
resonances ( or dynamically generated from other components than those
taken into account).  In this sense, in \cite{albertopheno} this idea
was put to work taking  experimental amplitudes for $\pi N$ above 1600 MeV,
where the chiral theory \cite{inoue}
   fails to deal with the $N^*(1650)$
resonance.

  Many baryon resonances are found in \cite{albertopheno} with strangeness zero, but we are only interested here on the state made out from 
$K \bar{K} N$ and coupled channels.  We find that in spite of the adding
new channels into the approach, the results of \cite{jidoenyo} remain and
a resonance appears around the energy predicted in \cite{jidoenyo}. There
is only a small novelty, the $K \bar{K}$ clusters in our 
approach both
around the $f_0(980)$ and the $a_0(980)$, while in \cite{jidoenyo} it was
mostly a $a_0(980) N$ state.

    It is interesting to note that such a state around 1920 MeV, could
have already been seen. This is the idea behind the work of 
\cite{albertoulf} which claims that this state could correspond to the
peak seen in the $\gamma p \to K^+ \Lambda$ reaction as we discuss in the next subsection.

\subsection{ Comparison of the $\gamma p \to K^+ \Lambda$ and
  $\gamma p \to K^+ \Sigma^0$ reactions}
  
      A peak of moderate strength on top of a large background is clearly seen for the $\gamma p \to K^+ \Lambda$ reaction around $1920$ MeV at all angles (see Fig. 18 of \cite{jefflab}). One can induce qualitatively a width for this peak of about 100 MeV or less. On the other hand, if one looks at the $\gamma p \to K^+ \Sigma^0$ reaction, one finds that starting from threshold a big large and broad structure develops, also visible at all angles (see Fig. 19 of \cite{jefflab}). The width of this structure is about 200-300 MeV. One can argue qualitatively that the relatively narrow peak of the
$\gamma p \to K^+ \Lambda$ reaction around 1920 MeV on top of a large background has nothing to do with the broad structure of $\gamma p \to K^+ \Sigma^0$ around 1900 MeV. A more quantitative argument can be provided by recalling that in \cite{lee} the broad structure of the 
$\gamma p \to K^+ \Sigma^0$ is associated to two broad $\Delta$ resonances in their model, which obviously can not produce any peak in the   $\gamma p \to K^+ \Lambda$ reaction, which filters isospin $1/2$ in the final state. Certainly, part of the structure is background, which is already obtained in chiral unitary theories at the low energies of the reaction  \cite{Borasoy:2007ku}.
   
\begin{figure*}[!htb]
\centering
\includegraphics[width=\textwidth]{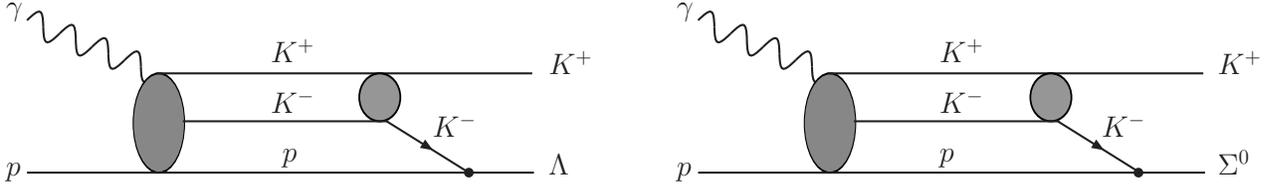}
\caption{Diagrams depicting the $\gamma p\rightarrow K^+\Lambda$, $\gamma p\rightarrow K^+\Sigma^0$ processes through the $1/2^+$ $N^*$ $K^+K^-N$ resonance of \cite{albertopheno,jidoenyo}.}\label{gpKKL}
\end{figure*}

    We thus adopt the position that the peak in the  $\gamma p \to K^+ \Lambda$ reaction is a genuine isospin $1/2$ contribution which does not show up in the $\gamma p \to K^+ \Sigma^0$ reaction. This feature would find a natural interpretation in the
picture of the state proposed in \cite{jidoenyo,albertopheno}. Indeed, in
those works the state at 1920 MeV is a $K \bar{K} N $ system in relative 
s-waves for all pairs.  The $\gamma p \to K^+ \Lambda$ and 
$\gamma p \to K^+ \Sigma^0$ reactions proceeding through the excitation of 
this resonance are depicted
in Fig.~\ref{gpKKL}. The two reactions are identical in this picture,
the only difference being the Yukawa coupling of the $K^-$ to the proton to
generate either a $\Lambda$ or a $\Sigma^0$.

The Yukawa couplings in SU(3) are well known and given in terms of the F and D
coefficients  \cite{Bernard:1995dp}, with $D+F=1.26$ and $D-F= 0.33$ 
\cite{Close:1993mv,Borasoy:2007ku}.  The particular couplings for 
$K^- p\rightarrow \Lambda$ and $K^-p\rightarrow\Sigma^0$ are e.g. given in~\cite{phimed}.

\begin{eqnarray}
V_{K^- p\rightarrow\Lambda}&=&-\frac{2}{\sqrt{3}}\frac{D+F}{2f}+\frac{1}{\sqrt{3}}\frac{D-F}{2f}\nonumber\\
\\
V_{K^- p\rightarrow\Sigma^0}&=&\frac{D-F}{2f}\nonumber
\end{eqnarray}

with $f$ the pion decay constant. Hence, the couplings are proportional to $-1.26$
and $0.33$ for the $K^- p \to \Lambda$ and $K^- p \to \Sigma^0$ vertices,
respectively. Therefore, it is clear that in this picture the signal of the
resonance in the $\gamma p \to K^+ \Lambda$ reaction is far larger than in the
$\gamma p \to K^+ \Sigma^0$  one, by as much as an order of magnitude in the 
case that
the resonance and background contributions  sum incoherently. The $3/2^+$ resonance 
used in the analysis of the $\gamma p \to K^+ \Lambda$ reaction in \cite{Nikonov:2007br} 
is also used for the $\gamma p \to K^+ \Sigma^0$  reaction in that work and also give a smaller contribution in this latter case. In the picture
of \cite{jidoenyo,albertopheno} the $1/2^+$ resonance also appears in 
the $\gamma p \to K^+ \Sigma^0$ reaction but with a smaller intensity than in the 
$\gamma p \to K^+ \Lambda$ one, as we have mentioned. 
  
\subsection{Test with polarization experiments}

In case the $J^P=1/2^+$ assignment was correct, an easy test can be carried out to rule out the $3/2^+$ state. 
The experiment consists in performing the 
$\gamma p \to K^+ \Lambda$ reaction with a
circularly polarized photon with helicity~1, thus $S_z=1$ with the $z$-axis
defined along the photon direction, together with a polarized proton of the
target with $S_z=1/2$ along the same direction. With this set up, the total spin
has $S^{tot}_z=3/2$. Since $L_z$ is zero with that choice of the $z$ direction,
then $J^{tot}_z=3/2$ and $J$ must be equal or bigger than $3/2$. Should the
resonant state be $J^P=1/2^+$, the peak signal would disappear for this polarization
selection, while it would remain if the resonance was a $J^P=3/2^+$ state. 
Thus, the disappearance of the signal with this polarization set up would rule out the 
$J^P=3/2^+$ assignment.

  Such type of polarization set ups have been done and are common in facilities
like ELSA at Bonn, MAMI B at Mainz or CEBAF at Jefferson Lab, where spin-3/2
and 1/2 cross sections, which play a crucial role
in the GDH sum rule, see e.g. Ref.~ \cite{Drechsel:1995az}, were measured in the
two-pion photoproduction \cite{Ahrens:2001qt,Ahrens:2007zzj} reaction.
The theoretical analysis of \cite{Nacher:2001yr} shows indeed that the
separation of the amplitudes in the spin channels provides information on the
resonances excited in the reaction.

Another test that can be carried out is by looking at the threshold
behavior of the cross section for the $\gamma p \to p K^+ K^-$ reaction. 
As a consequence of the existence of an s-wave resonance below 
threshold one finds an enhancement of the cross section around threshold
for this reaction, as well as an enhancement of the invariant mass of 
$K^+ K^-$ close to the sum of the masses of the two kaons, which results from
the strong coupling of these two particles to the $f_0(980)$ and
$a_0(980)$ resonances. An experiment on this reaction is already under way
and preliminary results already exist at LEPS \cite{nakaprivate}.

\section{X(2175) as a resonant state of the $\phi K \bar{K}$ system}

The discovery of the X(2175) $1^{--}$ resonance in $e^+ e^- \to \phi f_0(980)$ with
initial state radiation at BABAR \cite{Aubert:2006bu,Aubert:2007ur}, also
confirmed at BES in  $J/\Psi \to \eta \phi f_0(980)$ \cite{:2007yt}, has
stimulated research around its nontrivial nature  in terms of quark components. 
The possibility of it being a  tetraquark $s\bar{s} s \bar{s}$ is
investigated within QCD sum rules in \cite{Wang:2006ri}, and  as a 
 gluon hybrid $s \bar{s} g$ state has been discussed in 
\cite{Ding:2007pc,Close:2007ny}. A recent review on this issue can be seen in 
\cite{Zhu:2007wz}, where the basic problem of the expected large decay 
widths into two mesons of the states of these models, contrary to what 
is experimentally observed, 
is discussed. The basic data on this resonance from 
\cite{Aubert:2006bu,Aubert:2007ur} are $M_X = 2175 \pm 10$ MeV and 
$ \Gamma = 58 \pm 16 \pm 20$ MeV,
which are consistent with the numbers quoted from BES 
$M_X = 2186 \pm 10 \pm 6$ MeV and $ \Gamma = 65 \pm 25 \pm 17$ MeV. In 
Ref. \cite{Aubert:2007ur} an indication of this resonance is seen as an 
increase of the $K^+ K^- K^+ K^-$ cross
section around $2150$ MeV. A detailed theoretical study of the 
$e^+ e^- \to \phi f_0(980)$ reaction was done in Ref. \cite{Napsuciale:2007wp} by
means of loop diagrams involving kaons and $K^*$, using chiral amplitudes for the
$K \bar{K} \to \pi \pi$ channel which contains the $f_0(980)$ pole 
generated dynamically by the theory. 
The study revealed that the loop mechanisms reproduced the background but failed to
produce the peak around $ 2175$ MeV, thus reinforcing the claims for a new
resonance around this mass.

  In \cite{MartinezTorres:2008gy} a very different picture for the X(2175) resonance was advocated which allows for a reliable calculation and leads naturally to a very narrow 
width and no coupling to two pseudoscalar mesons.  The picture is that the  X(2175) is 
an ordinary resonant state of $\phi f_0(980)$ due to the interaction of these
components. The $f_0(980)$ resonance is dynamically generated from
the interaction of $\pi \pi$ and $K \bar{K}$ treated as coupled channels within the
chiral unitary approach of \cite{npa,kaiser,markushin}, qualifying 
as a kind of molecule with $\pi \pi$ and $K \bar{K}$ as its components, with a large coupling to 
$K \bar{K}$ and a weaker one to $\pi \pi$ [hence, the small width compared to that of the $\sigma(600)$]. 
Similar studies for the vector-pseudoscalar interaction have also been carried out using chiral dynamics 
in \cite{lutz,Roca:2005nm}, which lead to the dynamical generation of the low-lying axial vectors. In \cite{MartinezTorres:2008gy} the approach of Ref. \cite{Roca:2005nm} to deal with this part of the problem was used and 
 the $\phi K$ and $\phi \pi$ amplitudes were obtained in that approach.

To study
the  $\phi f_0(980)$ interaction, one is thus forced to investigate the three-body system 
$\phi K \bar{K}$ considering the interaction of the three components among
themselves and keeping in mind the expected strong correlations of the $K
\bar{K}$  system to make the $f_0(980)$ resonance. For this purpose one
solves the Faddeev equations  with coupled channels $\phi K^+ K^-$ and 
$\phi K^0 \bar{K^0}$. The picture is later complemented with the addition 
of the $\phi \pi \pi$ state  as a coupled channel. 
 The study benefits from  previous ones on the
$\pi\bar{K} N$ and $\pi\pi N$ along with their coupled channels done 
in \cite{MartinezTorres:2007sr,Khemchandani:2008rk}, where many 
$1/2^+$, strange, and nonstrange
low-lying baryon resonances of the Particle Data Group \cite{pdg}
were reproduced. This success encouraged us to extend the model of 
Refs. \cite{MartinezTorres:2007sr,Khemchandani:2008rk} to study the 
three-meson system, i.e., $\phi K\bar{K}$. 

\begin{figure*}[!htb]
\centering
\includegraphics[scale=1.1]{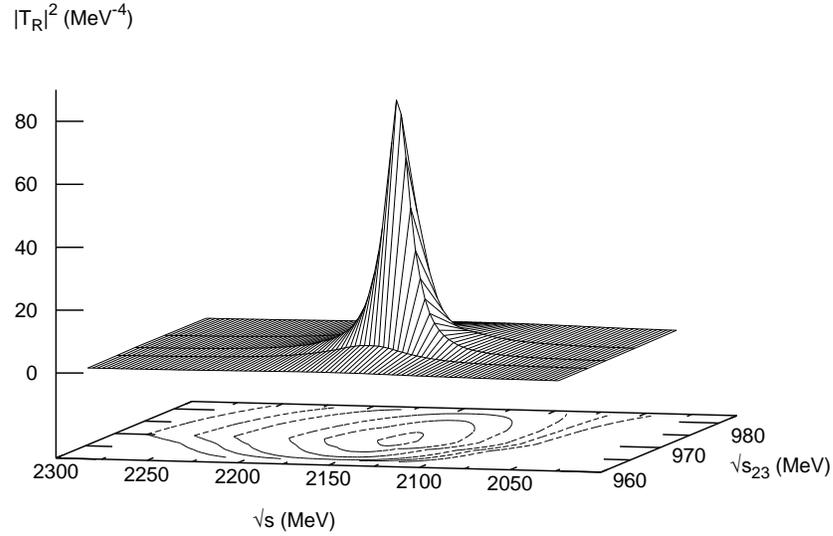}
\caption{The $\phi K \bar{K}$ squared amplitude in the isospin 0 configuration.}
\label{ampsq}
\end{figure*}

In fig. \ref{ampsq} we show $|T|^2$ as a function of the total energy and
the invariant mass of $K \bar{K}$.  We find a clear peak around 2150 MeV
for the total energy and 980 MeV for the invariant mass of $K \bar{K}$,
indicating that we have a resonance made basically from $\phi f_0(980)$,
which provides a natural explanation to the experimental features
observed for this resonance.

\section{The Y(4260) as a $J/\psi K \bar{K}$ system}
An enhancement in the data for the $\pi^+\pi^- J/\psi$ invariant mass spectrum 
was found near 4.26 GeV by the BABAR collaboration in a study of the 
$e^+ e^- \rightarrow \gamma_{ISR} \pi^+\pi^- J/\psi$ process\cite{babar1}. 
A fit to this data set was made by assuming a resonance with 4.26 GeV of mass 
and 50 to 90 MeV of width \cite{babar1}. The resonance was named as the $Y(4260)$ and it was found to be of $J^{PC} =1^{--}$ nature. Later on, an accumulation of events with similar characteristics in  the $\pi^+\pi^- J/\psi$, $\pi^0 \pi^0 J/\psi$ and the $K^+K^- J/\psi$ mass spectra was reported by the CLEO collaboration \cite{cleo1,cleo2}, thus confirming the results from BABAR. Following these works, the BELLE collaboration obtained the cross sections for the 
$e^+ e^- \rightarrow  \pi^+\pi^- J/\psi$ reaction in the 3.8 to 5.5 GeV region 
\cite{belle1}, by keeping all the interactions in the final state in $S$-wave, 
and found a peak at 4.26 GeV and a bump around 4.05 GeV.

Although the $Y(4260)$ does not seem to fit in to the charmonium spectrum  of 
the particle data group \cite{pdg} known up to $\sim$ 4.4 GeV, a proposal to 
accommodate it as a $4s$ state has been made in \cite{felipe}. Several other 
suggestions have been made for the interpretation of this state, for example, 
the authors of \cite{tetraquark} propose it to be a tetra-quark state, others 
propose a hadronic molecule of $D_{1} D$, $D_{0} D^*$ \cite{Ding,Albuquerque}, 
$\chi_{c1} \omega$ \cite{Yuan}, $\chi_{c1} \rho$ \cite{liu} and yet another idea is that it is a hybrid charmonium \cite{zhu} or charm baryonium \cite{Qiao}, etc. Within the available experimental information, none of these suggestions can be completely ruled out and its not clear if $Y(4260)$ possesses any of these structures dominantly or is a mixture of all of them. In Refs. \cite{eef1,eef2,eef3} the authors call the attention of the readers to the presence of the opening of the $D^*_s \bar{D}^*_s$ channel very close to the peak position of the $Y(4260)$ in the updated data from BABAR
\cite{babarupdate} and associate the peak corresponding to $Y(4260)$ to a 
$D^*_s \bar{D}^*_s$ cusp. A fit to the data from \cite{babar1,babarupdate} has
 been made in \cite{eef2} and the presence of a rather  broad bump around 
 4.35 GeV has been proposed.

\begin{figure*}[!htb]
\centering
\includegraphics[width=12cm]{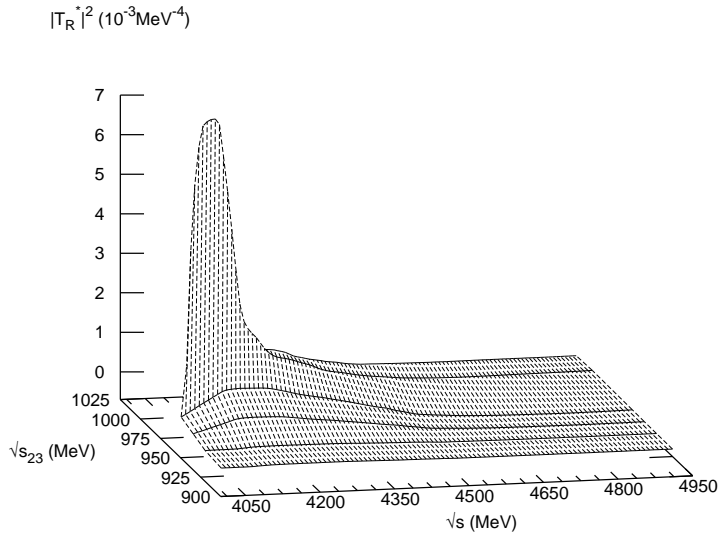}
\caption{$|T^*_{R}|^2$ for the $J/\psi K\bar{K}$ system in total 
isospin $I=0$ as a function of the total energy, $\sqrt{s}$, and the 
invariant mass of the $K\bar{K}$ subsystem, $\sqrt{s_{23}}$.}
\label{4260}
\end{figure*}

There are some peculiarities in the experimental findings which motivate us to 
carry out a study of the $J/\psi \pi \pi$ system. There is no enhancement found 
around 4.26 GeV in the process with the $D^* \bar{D}^*$ \cite{no4260} or other 
hadron final states \cite{pdg1} and it is concluded that $Y(4260)$ has an 
unusually strong coupling to the $\pi \pi J/\psi$ final state 
\cite{babar1,cleo1,cleo2,belle1}. Further, the data on the invariant mass of 
the $\pi \pi$ subsystem obtained by the Belle collaboration  \cite{belle1}, for 
total energy range, 3.8-4.2 GeV, 4.2-4.4 GeV and 4.4-4.6 GeV, have curious 
features. The $\pi \pi$ mass distribution data in 3.8-4.2 GeV and 4.4-4.6 GeV 
seem to follow the phase space, however, that corresponding to the 4.2-4.4 GeV 
total energy differs significantly from the phase space and shows an 
enhancement near $m_{\pi \pi} =$ 1 GeV. Do these findings indicate that the 
$Y(4260)$ has a strong coupling to $f_0(980) J/\psi$, similar to the $X(2175)$ 
to the $\phi f_{0}(980)$ \cite{babar2,bes}. It is interesting to recall that 
the $X(2175)$ was found as a dynamically generated resonance in the 
$\phi K \bar{K}$ system \cite{MartinezTorres:2008gy,alvarez} with the $K \bar{K}$ subsystem 
possessing the characteristics of the $f_0(980)$. Similarly, the $Y(4660)$ 
\cite{belle2} has been suggested as a $\psi^\prime f_0(980)$ resonance \cite{guo}. 
In order to find an answer to this question, in \cite{MartinezTorres:2009xb} the Faddeev 
equations were solved for the $J/\psi \pi \pi$  and $J/\psi K \bar{K}$ coupled channels 
and the results are shown in fig. \ref{4260}, where one sees that a peak of the amplitude squared is generated around an energy of 4150 MeV and $\sqrt s_{23}$ around 980 MeV.

\section{Conclusions}  
There have been several topics addressed in this talk around kaonic
clusters, some of them involving kaons and nucleons or kaons and vector
mesons. The conclusions have been appearing in each of the sections. As a
brief summary we can conclude that antikaons are very peculiar particles, 
with very strong interaction with nucleons and mesons, either pseudoscalar
or vectors.  They provide a glue to bind many systems, however, some of
them, particularly those involving several nucleons, seem to decay 
faster than 
we could observe.  Yet, we found some other systems which are long lived
enough, or sufficiently separated from other related states, such
that their observation is possible and many of them have already been
observed.  We also think this is just the beginning in the search for
these kind of new states and anticipate that many more states, of three 
body nature, or even with more particles, will eventually be seen in the
future. Devoting work to this issue at the present time seems to us a very
good scientific investment.


\begin{thebibliography}{99}

\bibitem{Eck95}
  G.~Ecker,
  Prog.\ Part.\ Nucl.\ Phys.\  {\bf 35}, 1 (1995)
  [arXiv:hep-ph/9501357].
  
\bibitem{Be95}
  V.~Bernard, N.~Kaiser and U.~G.~Meissner,
  Int.\ J.\ Mod.\ Phys.\  E {\bf 4}, 193 (1995)
  [arXiv:hep-ph/9501384].
  
\bibitem{angels}
  E.~Oset and A.~Ramos,
  Nucl.\ Phys.\  A {\bf 635}, 99 (1998)
  [arXiv:nucl-th/9711022].
  
\bibitem{npa}
  J.~A.~Oller and E.~Oset,
  Nucl.\ Phys.\  A {\bf 620} 438 (1997) 
  [Erratum-ibid.\  A {\bf 652} 407 (1999)].
  


\bibitem{juanoset}
  D.~Gamermann, J.~Nieves, E.~Oset and E.~R.~Arriola,
  arXiv:0911.4407 [hep-ph].
  
  

\bibitem{Koch94} V. Koch, Phys. Lett. B337 (1994) 7
\bibitem{WKW96} T. Waas, N. Kaiser and W. Weise, Phys. Lett.
B365 (1996) 12; {\it ibid.} B379 (1996) 34
\bibitem{Waas97} T. Waas and W.
Weise, Nucl. Phys. A625 (1997) 287
\bibitem{Lutz98} M. Lutz, Phys. Lett. B426 (1998) 12

\bibitem{angelsself}
  A.~Ramos and E.~Oset,
  Nucl.\ Phys.\ A {\bf 671} (2000) 481
  [arXiv:nucl-th/9906016].

\bibitem{schaffner}
  J.~Schaffner-Bielich, V.~Koch and M.~Effenberger,
  Nucl.\ Phys.\ A {\bf 669} (2000) 153
  [arXiv:nucl-th/9907095].

\bibitem{galself}
  A.~Cieply, E.~Friedman, A.~Gal and J.~Mares,
  Nucl.\ Phys.\ A {\bf 696} (2001) 173
  [arXiv:nucl-th/0104087].

\bibitem{Tolos:2006ny}
  L.~Tolos, A.~Ramos and E.~Oset,
  Phys.\ Rev.\  C {\bf 74} (2006) 015203
  [arXiv:nucl-th/0603033].
  
  
  \bibitem{okumura}
  S.~Hirenzaki, Y.~Okumura, H.~Toki, E.~Oset and A.~Ramos,
  Phys.\ Rev.\ C {\bf 61} (2000) 055205.


\bibitem{baca}
A.~Baca, C.~Garcia-Recio and J.~Nieves,
Nucl.\ Phys.\ A {\bf 673} (2000) 335
[arXiv:nucl-th/0001060].

  
\bibitem{hayano}
  R.~S.~Hayano {\it et al.}  [KEK-E570 Collaboration],
  Mod.\ Phys.\ Lett.\  A {\bf 23}, 2505 (2008).
  
\bibitem{akaishi:2002bg}
  Y.~Akaishi and T.~Yamazaki,
  Phys.\ Rev.\  C {\bf 65} (2002) 044005.
  
  
\bibitem{akainew}
  Y.~Akaishi, A.~Dote and T.~Yamazaki,
  Phys.\ Lett.\  B {\bf 613} (2005) 140
  [arXiv:nucl-th/0501040].
   
\bibitem{friedman-gal}
E. Friedman, A. Gal, and C. J. Batty, Nucl. Phys. A {\bf 579} (1994) 518 

\bibitem{Kaplan:1986yq}
  D.~B.~Kaplan and A.~E.~Nelson,
  Phys.\ Lett.\  B {\bf 175} (1986) 57.
   
 \bibitem{gal1}
  E.~Friedman, A.~Gal and J.~Mares,
  Phys.\ Rev.\  C {\bf 60} (1999) 024314
  [arXiv:nucl-th/9804072].

\bibitem{gal2}
  J.~Mares, E.~Friedman and A.~Gal,
  Phys.\ Lett.\  B {\bf 606} (2005) 295
  [arXiv:nucl-th/0407063].

\bibitem{gal3}
  J.~Mares, E.~Friedman and A.~Gal,
  Nucl.\ Phys.\  A {\bf 770} (2006) 84
  [arXiv:nucl-th/0601009].


\bibitem{gal4}
  D.~Gazda, E.~Friedman, A.~Gal and J.~Mares,
  Phys.\ Rev.\  C {\bf 76} (2007) 055204
  [arXiv:0708.2157 [nucl-th]].

 
  
\bibitem{gal}
  E.~Friedman and A.~Gal,
  Phys.\ Rept.\  {\bf 452} (2007) 89
  [arXiv:0705.3965 [nucl-th]].

  


\bibitem{muto}
  T.~Muto, T.~Maruyama and T.~Tatsumi,
  Phys.\ Rev.\  C {\bf 79} (2009) 035207
  [arXiv:0812.2537 [nucl-th]].


\bibitem{amigo1}
  T.~Maruyama, T.~Tatsumi, T.~Endo and S.~Chiba,
  Recent Res. Devel. Physics 7 (2006) 1 [arXiv:nucl-th/0605075].

\bibitem{amigo2}
  T.~Maruyama, T.~Tatsumi, D.~N.~Voskresensky, T.~Tanigawa, T.~Endo and S.~Chiba,
  Phys.\ Rev.\  C {\bf 73} (2006) 035802
  [arXiv:nucl-th/0505063].
  
\bibitem{Zhong:2004wa}
  X.~H.~Zhong, L.~Li, C.~H.~Cai and P.~Z.~Ning,
  Commun.\ Theor.\ Phys.\  {\bf 41} (2004) 573.
 
  
\bibitem{Zhong:2006hd}
  X.~H.~Zhong, G.~X.~Peng, L.~Li and P.~Z.~Ning,
  Phys.\ Rev.\  C {\bf 74} (2006) 034321.
  
\bibitem{Dang:2007ai}
  L.~Dang, L.~Li, X.~H.~Zhong and P.~Z.~Ning,
  Phys.\ Rev.\  C {\bf 75} (2007) 068201
  [arXiv:nucl-th/0701060].
  
  
 
\bibitem{toki}
  E.~Oset and H.~Toki,
  Phys.\ Rev.\  C {\bf 74} (2006) 015207
  [arXiv:nucl-th/0509048].
  

\bibitem{Hyodo:2007jq}
  T.~Hyodo and W.~Weise,
  Phys.\ Rev.\  C {\bf 77}, 035204 (2008)
  [arXiv:0712.1613 [nucl-th]].
  

\bibitem{akanuc}
  T.~Yamazaki and Y.~Akaishi,
  Nucl.\ Phys.\  A {\bf 792} (2007) 229
  [arXiv:nucl-ex/0609041].
  


  

   
\bibitem{Oset:2007vu}
  E.~Oset, V.~K.~Magas, A.~Ramos and H.~Toki,
proceedings of the IX International Conference on Hypernuclear and
Strange Particle Physics, Mainz (Germany), October 10-14, 2006. Edited by J.
Pochodzalla and Th. Walcher,
(Springer, Germany, 2007), 263

 
\bibitem{npangels}
  A.~Ramos, V.~K.~Magas, E.~Oset and H.~Toki,
  Nucl.\ Phys.\  A {\bf 804} (2008) 219.
  

\bibitem{shevchenko}
 N.~V.~Shevchenko, A.~Gal, J.~MaresJose Goity wrote: and J.~Revai,
 Phys.\ Rev.\  C {\bf 76}, 044004 (2007)
 [arXiv:0706.4393 [nucl-th]].

\bibitem{hyodo}
 A.~Dote, T.~Hyodo and W.~Weise,
 Nucl.\ Phys.\  A {\bf 804}, 197 (2008)
 [arXiv:0802.0238 [nucl-th]].

\bibitem{sato}
 Y.~Ikeda and T.~Sato,
 Phys.\ Rev.\  C {\bf 76}, 035203 (2007)
 [arXiv:0704.1978 [nucl-th]].

\bibitem{akaishi:2007cs}
 T.~Yamazaki and Y.~Akaishi,
 Phys.\ Rev.\  C {\bf 76}, 045201 (2007)
 [arXiv:0709.0630 [nucl-th]].
 
 
\bibitem{Suzuki:2004ep}
  T.~Suzuki {\it et al.},
  Phys.\ Lett.\  B {\bf 597} (2004) 263.
  
\bibitem{Agnello:2005qj}
  M.~Agnello {\it et al.}  [FINUDA Collaboration],
  Phys.\ Rev.\ Lett.\  {\bf 94} (2005) 212303.
  
\bibitem{:2007ph}
  M.~Agnello {\it et al.}  [FINUDA Collaboration],
  Phys.\ Lett.\  B {\bf 654} (2007) 80
  [arXiv:0708.3614 [nucl-ex]].
  
\bibitem{Magas:2006fn}
  V.~K.~Magas, E.~Oset, A.~Ramos and H.~Toki,
  Phys.\ Rev.\  C {\bf 74} (2006) 025206
  [arXiv:nucl-th/0601013];

\bibitem{Crimea}
  V.~K.~Magas, E.~Oset, A.~Ramos and H.~Toki,
  [arXiv:nucl-th/0611098].
  


 
 
 
 
\bibitem{Magas:2008bp}
  V.~K.~Magas, E.~Oset and A.~Ramos,
  Phys.\ Rev.\  C {\bf 77}, 065210 (2008)
  [arXiv:0801.4504 [nucl-th]];

 [arXiv:0901.1086 [nucl-th]];


  
\bibitem{Suzuki:2007kn}
  T.~Suzuki {\it et al.}  [KEK-PS E549 Collaboration],
  Phys.\ Rev.\  C {\bf 76} (2007) 068202
  [arXiv:0709.0996 [nucl-ex]].



\bibitem{Kishimoto:2007zz}
  T.~Kishimoto {\it et al.},
  Prog.\ Theor.\ Phys.\  {\bf 118}, 181 (2007).
  
\bibitem{magasflight}
V.~K.~Magas, J. Yamagata-Sekihara, S. Hirenzaki, E. Oset,  A. Ramos,
  arXiv:0911.3614 [nucl-th];
%
  arXiv:0911.2091 [nucl-th];
%
A.~Ramos, V.~K.~Magas, J. Yamagata-Sekihara, S. Hirenzaki, E. Oset,
  arXiv:0911.4841 [nucl-th].



\bibitem{Morimatsu:1994sx}
  O.~Morimatsu and K.~Yazaki,
  Prog.\ Part.\ Nucl.\ Phys.\  {\bf 33}, 679 (1994).
  
\bibitem{YH}
 J. Yamagata and S. Hirenzaki, 
Eur. Phys. J. {\bf A31}, 255 (2007).

\bibitem{Yamagata:2007cp}
  J.~Yamagata, H.~Nagahiro, R.~Kimura and S.~Hirenzaki,
  Phys.\ Rev.\  C {\bf 76}, 045204 (2007)
  [arXiv:0708.0459 [nucl-th]].
  
  
  
\bibitem{YamagataSekihara:2008ji}
  J.~Yamagata-Sekihara, D.~Jido, H.~Nagahiro and S.~Hirenzaki,
  Phys.\ Rev.\  C {\bf 80}, 045204 (2009)
  [arXiv:0812.4359 [nucl-th]].
  
  
\bibitem{Katz:1970ng}
  P.~A.~Katz, K.~Bunnell, M.~Derrick, T.~Fields, L.~G.~Hyman and G.~Keyes,
  Phys.\ Rev.\  D {\bf 1} (1970) 1267.
   
\bibitem{Oset:1986yi}
  E.~Oset, Y.~Futami and H.~Toki,
  Nucl.\ Phys.\  A {\bf 448}, 597 (1986).
  
\bibitem{Weyer:1990ye}
  H.~J.~Weyer,
  Phys.\ Rept.\  {\bf 195} (1990) 295.
 
\bibitem{Salcedo:1987md}
  L.~L.~Salcedo, E.~Oset, M.~J.~Vicente-Vacas and C.~Garcia-Recio,
  Nucl.\ Phys.\  A {\bf 484}, 557 (1988).

\bibitem{Kosmas:1996fh}
  T.~S.~Kosmas and E.~Oset,
  Phys.\ Rev.\  C {\bf 53}, 1409 (1996).
  
\bibitem{Albertus:2001pb}
  C.~Albertus, J.~E.~Amaro and J.~Nieves,
  Phys.\ Rev.\ Lett.\  {\bf 89}, 032501 (2002)
  [arXiv:nucl-th/0110046].
  
\bibitem{Nieves:2004wx}
  J.~Nieves, J.~E.~Amaro and M.~Valverde,
  Phys.\ Rev.\  C {\bf 70}, 055503 (2004)
  [Erratum-ibid.\  C {\bf 72}, 019902 (2005)]
  [arXiv:nucl-th/0408005].
  

\bibitem{Ramos:2007zz}
  A.~Ramos, V.~K.~Magas, E.~Oset and H.~Toki,
  Eur.\ Phys.\ J.\  {\bf A31}, 684 (2007);
%
  [arXiv:nucl-th/0702019].

 \bibitem{DW-HYP06} 
A.~Dot\'e and W.~Weise, Prog. Theor. Phys. Suppl. {\bf 168}, 593 (2007);
A.~Dot\'e and W.~Weise, nucl-th/0701050, in: Proceedings HYP06, 
{\it ``Hypernuclear and Strange Particle Physics''}, J. Pochodzalla and Th. Walcher, ed., p 249, 
Springer, Heidelberg, 2007 (ISBN 978-3-540-76365-9). 

\bibitem{ppK_Lstar:Oka} A.~Arai, M.~Oka and S.~Yasui, 
\newblock Prog. Theor. Phys. {\bf 119}, 103 (2008). 



\bibitem{ppK_Reac:Koike07} T.~Koike and T.~Harada, 
Phys. Lett. {\bf B652}, 262 (2007); Nucl. Phys. {\bf A804}, 231 (2008).

\bibitem{Dalitz:1967fp}
  R.~H.~Dalitz, T.~C.~Wong and G.~Rajasekaran,
  Phys.\ Rev.\  {\bf 153} (1967) 1617.

\bibitem{Siegel:1988rq}
  P.~B.~Siegel and W.~Weise,
  Phys.\ Rev.\  C {\bf 38}, 2221 (1988).
  

 
\bibitem{Jido:2003cb}
  D.~Jido, J.~A.~Oller, E.~Oset, A.~Ramos and U.~G.~Meissner,
  Nucl.\ Phys.\  A {\bf 725}, 181 (2003)
  [arXiv:nucl-th/0303062].
  
\bibitem{Thomas:1973uh}
  D.~W.~Thomas, A.~Engler, H.~E.~Fisk and R.~W.~Kraemer,
  Nucl.\ Phys.\  B {\bf 56} (1973) 15.
 

\bibitem{mko}
  A.~Martinez Torres, K.~P.~Khemchandani and E.~Oset,
  Phys.\ Rev.\  C {\bf 77}, 042203 (2008) .
  
 

\bibitem{MartinezTorres:2008is}
 A.~Martinez Torres, K.~P.~Khemchandani and E.~Oset,
 Eur.\ Phys.\ J.\  A {\bf 35}, 295 (2008)
 [arXiv:0805.3641 [nucl-th]].
 
 
 
 
\bibitem{pdg}
  C.~Amsler {\it et al.}  [Particle Data Group],
  Phys.\ Lett.\  B {\bf 667}, 1 (2008).

\bibitem{jidoenyo}
 D.~Jido and Y.~Kanada-En'yo,
 Phys.\ Rev.\  C {\bf 78}, 035203 (2008)
 [arXiv:0806.3601 [nucl-th]].



\bibitem{albertopheno}
 A.~Martinez Torres, K.~P.~Khemchandani and E.~Oset,
 arXiv:0812.2235 [nucl-th].
 
 
\bibitem{inoue}
  T.~Inoue, E.~Oset and M.~J.~Vicente Vacas,
  Phys.\ Rev.\  C {\bf 65}, 035204 (2002)
  [arXiv:hep-ph/0110333].
  
  
\bibitem{albertoulf}
  A.~Martinez Torres, K.~P.~Khemchandani, U.~G.~Meissner and E.~Oset,
  Eur.\ Phys.\ J.\  A {\bf 41}, 361 (2009)
  [arXiv:0902.3633 [nucl-th]].
  
  
  
\bibitem{jefflab}
 R.~Bradford {\it et al.}  [CLAS Collaboration],
 Phys.\ Rev.\  C {\bf 73}, 035202 (2006)
 [arXiv:nucl-ex/0509033].


  \bibitem{lee}
  F.~X.~Lee, T.~Mart, C.~Bennhold and L.~E.~Wright,
  Nucl.\ Phys.\  A {\bf 695}, 237 (2001)
  [arXiv:nucl-th/9907119].

  


\bibitem{Borasoy:2007ku}
  B.~Borasoy, P.~C.~Bruns, U.~G.~Meissner and R.~Nissler,
  Eur.\ Phys.\ J.\  A {\bf 34}, 161 (2007)
  [arXiv:0709.3181 [nucl-th]].
  
  
\bibitem{Bernard:1995dp}
 V.~Bernard, N.~Kaiser and U.-G.~Mei{\ss}ner,
 Int.\ J.\ Mod.\ Phys.\  E {\bf 4}, 193 (1995)
 [arXiv:hep-ph/9501384].
 
 
\bibitem{Close:1993mv}
 F.~E.~Close and R.~G.~Roberts,
 Phys.\ Lett.\  B {\bf 316}, 165 (1993)
 [arXiv:hep-ph/9306289].
 
 


\bibitem{phimed}
 E.~Oset and A.~Ramos,
 Nucl.\ Phys.\  A {\bf 679}, 616 (2001)
 [arXiv:nucl-th/0005046].


 


 

\bibitem{Nikonov:2007br}
 V.~A.~Nikonov, A.~V.~Anisovich, E.~Klempt, A.~V.~Sarantsev and U.~Thoma,
 Phys.\ Lett.\  B {\bf 662}, 245 (2008)
 [arXiv:0707.3600 [hep-ph]].






\bibitem{Drechsel:1995az}
 D.~Drechsel,
 Prog.\ Part.\ Nucl.\ Phys.\  {\bf 34}, 181 (1995)
 [arXiv:nucl-th/9411034].
 
\bibitem{Ahrens:2001qt}
 J.~Ahrens {\it et al.}  [GDH Collaboration and A2 Collaboration],
 Phys.\ Rev.\ Lett.\  {\bf 87}, 022003 (2001)
 [arXiv:hep-ex/0105089].
\bibitem{Ahrens:2007zzj}
 J.~Ahrens {\it et al.}  [GDH and A2 Collaborations],
 Eur.\ Phys.\ J.\  A {\bf 34}, 11 (2007).
 
\bibitem{Nacher:2001yr}
 J.~C.~Nacher and E.~Oset,
 Nucl.\ Phys.\  A {\bf 697}, 372 (2002)
 [arXiv:nucl-th/0106005].


\bibitem{nakaprivate} T. Nakano, private communication.
 



\bibitem{Aubert:2006bu}
  B.~Aubert {\it et al.}
  Phys.\ Rev.\  D {\bf 74} 091103 (2006).
  
\bibitem{Aubert:2007ur}
  B.~Aubert {\it et al.}
  Phys.\ Rev.\  D {\bf 76} 012008 (2007).
  
\bibitem{:2007yt}
 M.~Ablikim {\it et al.}  [BES Collaboration],
  Phys.\ Rev.\ Lett.\  {\bf 100}, 102003 (2008).

 
\bibitem{Wang:2006ri}
  Z.~G.~Wang,
  Nucl.\ Phys.\  A {\bf 791} 106 (2007).
  
\bibitem{Ding:2007pc}
  G.~J.~Ding and M.~L.~Yan,
  Phys.\ Lett.\  B {\bf 657} 49 (2007).
  
\bibitem{Close:2007ny}
  F.~E.~Close,
  arXiv:0706.2709 [hep-ph].
   
\bibitem{Zhu:2007wz}
  S.~L.~Zhu,
  Int.\ J.\ Mod.\ Phys.\  E {\bf 17}, 283 (2008).

\bibitem{Napsuciale:2007wp}
  Phys.\ Rev.\  D {\bf 76} 074012 (2007).
  
\bibitem{MartinezTorres:2008gy}
 A.~Martinez Torres, K.~P.~Khemchandani, L.~S.~Geng, M.~Napsuciale and E.~Oset,
 Phys.\ Rev.\  D {\bf 78}, 074031 (2008)
 [arXiv:0801.3635 [nucl-th]].



  
\bibitem{kaiser}
  N.~Kaiser,
  Eur.\ Phys.\ J.\  A {\bf 3} 307 (1998).
  
\bibitem{markushin}
 V.~E.~Markushin,
 Eur.\ Phys.\ J.\  A Meissner{\bf 8} 389 (2000).

\bibitem{lutz}
  M.~F.~M.~Lutz and E.~E.~Kolomeitsev,
  Nucl.\ Phys.\  A {\bf 730}, 392 (2004).

  
\bibitem{Roca:2005nm}
  L.~Roca, E.~Oset and J.~Singh,
  Phys.\ Rev.\  D {\bf 72} 014002 (2005).


\bibitem{MartinezTorres:2007sr}
  A.~Martinez Torres, K.~P.~Khemchandani and E.~Oset,
  Phys.\ Rev.\  C {\bf 77}, 042203 (2008).
 






\bibitem{Khemchandani:2008rk}
 K.~P.~Khemchandani, A.~Martinez Torres and E.~Oset,
 Eur.\ Phys.\ J.\  A {\bf 37}, 233 (2008)
 [arXiv:0804.4670 [nucl-th]].









  
\bibitem{babar1}
  B.~Aubert {\it et al.}  [BABAR Collaboration],
  Phys.\ Rev.\ Lett.\  {\bf 95} 142001(2005).


\bibitem{cleo1}
  T.~E.~Coan {\it et al.}  [CLEO Collaboration],
  Phys.\ Rev.\ Lett.\  {\bf 96}, 162003 (2006).

\bibitem{cleo2}
  Q.~He {\it et al.}  [CLEO Collaboration],
  Phys.\ Rev.\  D {\bf 74}, 091104 (2006).

\bibitem{belle1}
C.~Z.~Yuan {\it et al.}  [Belle Collaboration],
  Phys.\ Rev.\ Lett.\  {\bf 99}, 182004 (2007).



\bibitem{felipe}
  F.~J.~Llanes-Estrada,
  Phys.\ Rev.\  D {\bf 72}, 031503 (2005).

\bibitem{tetraquark}
  L.~Maiani, V.~Riquer, F.~Piccinini and A.~D.~Polosa,
  Phys.\ Rev.\  D {\bf 72}, 031502 (2005).


\bibitem{Ding}
  G.~J.~Ding,
  Phys.\ Rev.\  D {\bf 79}, 014001 (2009).



\bibitem{Albuquerque}
  R.~M.~Albuquerque and M.~Nielsen,
  Nucl.\ Phys.\  A {\bf 815}, 53 (2009).



\bibitem{Yuan}
  C.~Z.~Yuan, P.~Wang and X.~H.~Mo,
  Phys.\ Lett.\  B {\bf 634}, 399 (2006).

\bibitem{liu}
  X.~Liu, X.~Q.~Zeng and X.~Q.~Li,
  Phys.\ Rev.\  D {\bf 72}, 054023 (2005).



\bibitem{zhu}
  S.~L.~Zhu,
  Phys.\ Lett.\  B {\bf 625}, 212 (2005).


\bibitem{Qiao}
  C.~F.~Qiao,
  Phys.\ Lett.\  B {\bf 639}, 263 (2006).



\bibitem{eef1}
  E.~van Beveren and G.~Rupp,
  arXiv:hep-ph/0605317.



\bibitem{eef2}
  E.~van Beveren and G.~Rupp,
  arXiv:0904.4351 [hep-ph].



\bibitem{eef3}
  E.~van Beveren and G.~Rupp,
  arXiv:0905.1595 [hep-ph].




\bibitem{babarupdate}
  B.~Aubert {\it et al.}  [BaBar Collaboration],
  arXiv:0808.1543 [hep-ex].


\bibitem{no4260}
  B.~Aubert {\it et al.}  [BABAR Collaboration],
  arXiv:0903.1597 [hep-ex].



\bibitem{pdg1}
  S.~Eidelman {\it et al.}  [Particle Data Group],
  Phys.\ Lett.\  B {\bf 592}, 1 (2004).


\bibitem{babar2}
  B.~Aubert {\it et al.}  [BABAR Collaboration],
  Phys.\ Rev.\  D {\bf 74}, 091103 (2006).

\bibitem{bes}
  M.~Ablikim {\it et al.}  [BES Collaboration],
  Phys.\ Rev.\ Lett.\  {\bf 100}, 102003 (2008).



\bibitem{alvarez}
  L.~Alvarez-Ruso, J.~A.~Oller and J.~M.~Alarcon,
  Phys.\ Rev.\  D {\bf 80}, 054011 (2009)
  [arXiv:0906.0222 [hep-ph]].
  

\bibitem{belle2}
  X.~L.~Wang {\it et al.}  [Belle Collaboration],
  Phys.\ Rev.\ Lett.\  {\bf 99}, 142002 (2007) .
  
\bibitem{guo}
F.~K.~Guo, C.~Hanhart and U.~G.~Meissner,
  Phys.\ Lett.\  B {\bf 665}, 26 (2008).


\bibitem{MartinezTorres:2009xb}
  A.~Martinez Torres, K.~P.~Khemchandani, D.~Gamermann and E.~Oset,
  Phys.\ Rev.\  D {\bf 80}, 094012 (2009)
  [arXiv:0906.5333 [nucl-th]].








 
 



 
\end{thebibliography}
\end{document}